\documentclass[10pt,a4paper]{article}

\usepackage{fullpage}
\usepackage{amssymb,amsmath,mathrsfs}
\usepackage{tensor}

\newtheorem{definition}{Definition}[section]
\newtheorem{lemma}[definition]{Lemma}
\newtheorem{proposition}[definition]{Proposition}
\newtheorem{theorem}[definition]{Theorem}
\newtheorem{corollary}[definition]{Corollary}
\newtheorem{remark}[definition]{Remark}

\newenvironment{proof*}{\smallskip\par\noindent\emph{Proof. }
 \ignorespaces}{\hfill$\Box$\smallskip\par\ignorespaces}
\newenvironment{proofsketch*}{\smallskip\par\noindent
 \emph{Sketch of proof. }\ignorespaces}
 {\hfill$\oslash$\smallskip\par\ignorespaces}

\title{\textbf{Static symmetric solutions of the semi-classical Einstein-Klein-Gordon system}}
\author{Ko Sanders\thanks{E-mail:
jacobus.sanders@dcu.ie}}
\date{17 September 2021}

\begin{document}
\maketitle

\begin{abstract}
We consider solutions of the semi-classical Einstein-Klein-Gordon system with a cosmological constant $\Lambda\in\mathbb{R}$, where the spacetime is given by Einstein's static metric on $\mathbb{R}\times\mathbb{S}^3$ with a round sphere of radius $a>0$ and the state of the scalar quantum field has a two-point distribution $\omega_2$ that respects all the symmetries of the metric. We assume that the mass $m\ge0$ and scalar curvature coupling $\xi\in\mathbb{R}$ of the field satisfy $m^2+\xi R>0$, which entails the existence of a ground state. We do not require states to be Hadamard or quasi-free, but the quasi-free solutions are characterised in full detail.

The set of solutions of the semi-classical Einstein-Klein-Gordon system depends on the choice of the parameters $(a,\Lambda,m,\xi)$ and on the renormalisation constants in the renormalised stress tensor of the scalar field. We show that the set of solutions is either (i) the empty set, or (ii) the singleton set containing only the ground state, or (iii) a set with infinitely many elements. We characterise the ranges of the parameters and renormalisation constants where each of these alternatives occur. We also show that all quasi-free solutions are given by density matrices in the ground state representation and we show that in cases (ii) and (iii) there is a unique quasi-free solution which minimises the von Neumann entropy. When $m=0$ this unique state is a $\beta$-KMS state. We argue that all these conclusions remain valid in the reduced order formulation of the semi-classical Einstein equation.
\end{abstract}

\section{Introduction}

Although there is as yet no full theoretical description of quantum gravity, it is widely accepted that such a theory should admit a semi-classical limit, where the quantum aspects of the gravitational field become negligible. In this limit the theory is expected to be described by the equations of motion of the quantum fields and the semi-classical Einstein equation,\footnote{We use units in which $c=\hbar=1$ unless stated otherwise.}
\begin{equation}\label{Eqn_semicl}
\frac{1}{\kappa}(G_{ab}+\Lambda g_{ab})=\langle T^{\mathrm{ren}}_{ab}\rangle_{\omega}\,,
\end{equation}
where $G_{ab}=R_{ab}-\frac12Rg_{ab}$ is the Einstein tensor, $\Lambda$ is a cosmological constant, $\kappa=8\pi G_N$ a multiple of Newton's constant $G_N$ and $T^{\mathrm{ren}}_{ab}$ is the renormalised stress-energy-momentum tensor of the quantum matter, whose expectation value is taken in the state $\omega$.

The semi-classical Einstein equation is fraught with ambiguities and problems. For interacting quantum fields the interaction is usually treated perturbatively and the quantum stress tensor $T^{\mathrm{ren}}_{ab}$ is consequently given by a formal series expansion in the coupling constants \cite{Hollands+2005}. For free fields the quantum stress tensor can be defined rigorously, using a local and generally covariant renormalisation scheme, but even in this case there are ambiguities in its definition, parametrised by renormalisation constants. This also leads to ambiguities in the right-hand side of (\ref{Eqn_semicl}), which cannot be fixed without a full theory of quantum gravity or observational input. Furthermore, the right-hand side of (\ref{Eqn_semicl}) depends on a choice of state $\omega$ and one should couple this equation to the equations of motion of the quantum fields to solve for the metric and for the state $\omega$ together. The Cauchy problem for these coupled equations is difficult, because the required renormalisation of the stress tensor typically forces one to prescribe more time derivatives than one would expect classically. In general it is unknown how to give the coupled system a well-posed initial value formulation, even in the simplest toy model case of a single free real scalar quantum field $\phi$. A reduction of order scheme has been proposed to obtain a well-posed initial value formulation and to remove (some of the) spurious solutions \cite{Parker+1993}. For a discussion of the range of validity of the semi-classical Einstein equation and of the reduction of order scheme we refer to \cite{Flanagan+1996}.

Attempts to solve the semi-classical Einstein-Klein-Gordon system usually invoke a lot of symmetry and often some approximations to simplify the problem. The Minkowski vacuum of a free scalar field is a solution essentially by definition. Wald \cite{Wald1978} found additional solutions for conformally invariant quantum fields in a conformally flat spacetime of dimension two or four and in the conformal vacuum state. This analysis covers the cosmologically important case of Friedman-Lema\^{\i}tre-Robertson-Walker (FLRW) spacetimes, but only for conformally invariant fields. Most later works have not found a full solution for the state, focussing instead on the physically important effects on the metric on flat (i.e.\ $k=0$) FLRW spacetimes. This was done e.g.\ by using linear perturbations around a conformal vacuum \cite{Horowitz+1980}, by using numerical methods \cite{Anderson1986}, or by invoking some general properties of the state \cite{Dappiaggi+2008,Starobinsky1980,Suen1989}. Notable exceptions are the works of Pinamonti \cite{Pinamonti2011} and Pinamonti and Siemssen \cite{Pinamonti+2015} (see also \cite{Eltzner+2011}), who have shown the existence of solutions for a massive, conformally coupled scalar field on flat FLRW spacetimes by prescribing initial data at past null infinity or at a finite time. A recent extension of these results by Gottschalk and Siemssen \cite{Gottschalk+2018} (see also \cite{Meda+2020}) works for general curvature coupling and without restricting the renormalisation freedom. However, the solutions in these papers are found using Banach's fixed point theorem and they are not very explicit.

In this paper we will consider static solutions to the semi-classical Einstein-Klein-Gordon system. For many physical systems, static solutions are studied before the dynamics is included, because they are often simpler to handle and more easily accessible to observation. It appears, however, that the static solutions of the semi-classical Einstein-Klein-Gordon system have not been studied before (except the Minkowski vacuum). One reason may be that the system under consideration does not have any interesting classical static solutions. Nevertheless, for the semi-classical system we will find a class of rather explicit solutions by considering ultra-static spacetimes with maximally symmetric spatial slices. Since we will consider spatially compact spacetimes, these solutions may also serve as a first step towards investigations of closed FLRW spacetimes.

In particular, we require the spacetime to be given by Einstein's static universe
\begin{align}\label{Eqn_Estatic}
M=\mathbb{R}\times\mathbb{S}^3,&\qquad g=-\mathrm{d}t^2+a^2h\,,
\end{align}
with radius $a>0$, Killing time coordinate $t$ and $h$ the round metric on the unit three-sphere. We use the simple toy-model of an Einstein-Klein-Gordon system, which is described classically by the action functional
\begin{align}\label{Eqn_action}
S[g,\varphi]&=\frac{1}{2\kappa}\int_M\ R-2\Lambda\ \mathrm{dvol}_g\ -\ \frac12\int_M\ |\nabla\varphi|^2+m^2\varphi^2+\xi R\varphi^2\ \mathrm{dvol}_g \,.
\end{align}
Variation w.r.t.~the classical fields $\varphi$ and the inverse metric $g^{ab}$ yields, respectively, the Klein-Gordon and the Einstein equation,
\begin{align}
(-\Box+m^2+\xi R)\varphi&=0\,,\label{Eqn_clKG}\\
\frac{1}{\kappa}(G_{ab}+\Lambda g_{ab})&=T_{ab}\,,\label{Eqn_clEinstein}
\end{align}
where the stress tensor of the scalar field is given by
\begin{align}\label{Eqn_T}
T_{ab}&:=\nabla_{(a}\varphi\nabla_{b)}\varphi-\frac12g_{ab}(|\nabla\varphi|^2+m^2\varphi^2)+\xi(\varphi^2G_{ab}-\nabla_a\nabla_b\varphi^2+g_{ab}\Box\varphi^2)\,.
\end{align}
One verifies by direct computation using (\ref{Eqn_clKG}) that $\nabla^aT_{ab}=0$. For $m=0$ and $\xi=\xi_c=\frac16$ the field $\varphi$ is conformally invariant.

It is well-known how to quantise the field in a local and generally covariant way on every globally hyperbolic spacetime and how to find $T^{\mathrm{ren}}_{ab}$ and its renormalisation freedom from a Hadamard regularisation scheme \cite{WaldQFT,Brunetti+3,Hollands+2005}. Due to the symmetry of the Einstein static universe, the renormalisation freedom simplifies considerably, as we will see in Section \ref{Sec_Stress}. 

We will denote the quantum field by $\phi$, to distinguish it from the classical field $\varphi$. We will mostly focus on quasi-free (Gaussian) states, which are determined entirely in terms of their two-point distribution
\begin{align}\label{Eqn_def2pt}
\omega_2(x,x')&:= \langle \phi(x)\phi(x')\rangle_{\omega} \,,
\end{align}
which is assumed to be a distribution on $M\times M$. Its anti-symmetric part $\omega_{2-}(x,x'):=\frac12(\omega_2(x,x')-\omega_2(x',x))$ is completely fixed by the canonical commutation relations, but its symmetric part $\omega_{2+}(x,x'):=\frac12(\omega_2(x,x')+\omega_2(x',x))$ is only restricted by the requirements that $\omega_2$ should be of positive type and a solution to (\ref{Eqn_clKG}) in each argument. An especially nice class of states are the Hadamard states, which are characterised by their short distance singularity structure \cite{WaldQFT,Radzikowski1996} and which have a finite and smoothly varying expectation value of $T^{\mathrm{ren}}_{ab}$, but we will not insist that our solutions have the Hadamard property.

We will assume that the free scalar quantum field has a mass $m\ge0$ and a scalar curvature coupling $\xi\in\mathbb{R}$ such that $m^2+\xi R>0$. This entails that a ground state exists, as we will briefly review in Section \ref{Sec_Ground} below. In that section we will study the ground states in an Einstein static universe in some detail and Equation (\ref{Eqn_ground5}) gives an explicit expression for the two-point distribution in the conformally coupled case. In Section \ref{Sec_SymStates} we will analyse all quasi-free states that respect all the symmetries of the spacetime and we characterise their two-point distributions in terms of Gegenbauer polynomials on $\mathbb{S}^3$. This will allow us to determine the set of solutions to the semi-classical Einstein equation (\ref{Eqn_semicl}) rather explicitly. The renormalised stress tensor and its renormalisation freedom are discussed in Section \ref{Sec_Stress}. Section \ref{sec_SSE} contains our main results on the solutions of the semi-classical Einstein-Klein-Gordon system with a symmetric two-point distribution. We show in particular that the set of such solutions is either empty, the singleton set consisting of the ground state, or a set with infinitely many elements. If $m\not=0$ or $\Lambda\not=0$ all three cases can occur for suitable choices of the renormalisation constants. In Section \ref{sec_properties} we argue that all solutions with a vanishing one-point distribution also solve the reduced order form of the semi-classical Einstein equation. Moreover, we show that all quasi-free solutions are given by density matrices in the ground state representation and, if the system admits a solution, then there is a unique quasi-free solution which minimises the von Neumann entropy. When $m=0$ this unique solution is a $\beta$-KMS state. We conclude with a brief discussion in Section \ref{sec:discussion}.

\section{Ground states in Einstein's static universe}\label{Sec_Ground}

In this section we will consider the ground state of a non-minimally coupled scalar field in Einstein's static universe. Before we turn our attention to this specific spacetime, we will briefly review the construction and properties of ground and KMS states in a general stationary, globally hyperbolic spacetime $M$, i.e.~we assume that there is a complete time-like and future pointing Killing field $v^a$. Some of this material is taken from the review \cite{KS2013}.

Two-point distributions for the ground and thermal (KMS) states of a free scalar field can be constructed from the Hilbert space of (classical) finite energy solutions of the field. If $\varphi$ is a classical solution to the Klein-Gordon equation (\ref{Eqn_clKG}) with space-like compact support, then we can define its total energy by
\begin{align}\label{Eqn_totalE}
\mathcal{E}(\varphi)&:=\int_{\Sigma}n^av^bT_{ab}\,,
\end{align}
where $\Sigma\subset M$ is a smooth spacelike Cauchy surface with future normal vector field $n^a$. Because
$\nabla^av^bT_{ab}=v^b\nabla^aT_{ab}=0$, the total energy is independent of the choice of Cauchy surface, as may be seen using Stokes' theorem. When $\mathcal{E}(\varphi)>0$ for all non-zero $\varphi$, we can define a norm on the space of solutions $\varphi$ with space-like compact support by setting $\|\varphi\|:=\sqrt{\mathcal{E}(\varphi)}$. This norm comes from an inner product, which allows us to complete the space of solutions to the Hilbert space $\mathcal{H}_e$ of classical finite energy solutions.

We can also write the total energy as
\begin{align}\label{Eqn_totalE2}
\mathcal{E}(\varphi)&=\int_{\Sigma}n^av^b\tilde{T}_{ab}\,,
\end{align}
in terms of the modified (simplified) stress tensor $\tilde{T}_{ab}$, which is given by
\begin{align}\label{Eqn_simpleT}
\tilde{T}_{ab}&=\nabla_{(a}\varphi\nabla_{b)}\varphi-\frac12g_{ab}(|\nabla\varphi|^2+m^2\varphi^2+\xi R\varphi^2)\,.
\end{align}
The equality (\ref{Eqn_totalE2}) is shown in Appendix \ref{App_TotalE} and together with Equation (\ref{Eqn_simpleT}) it provides a simpler formula for the energy norm than Equations (\ref{Eqn_totalE}, \ref{Eqn_T}). Moreover, using the simplified stress tensor we see that $\mathcal{E}(\varphi)>0$ for all non-zero solutions $\varphi$ with space-like compact support as soon as $m^2+\xi R\ge0$ everywhere and either $\Sigma$ is non-compact or $m^2+\xi R>0$ somewhere. (Note that \cite{KS2013} also uses the simplified formula for the total energy, but without the justification that we provide in Appendix \ref{App_TotalE}.)

The Hilbert space $\mathcal{H}_e$ of classical finite energy solutions contains a dense subspace with elements of the form $E(f)$, where $f\in C_0^{\infty}(M)$ and $E=E^--E^+$ is the difference of the advanced and retarded fundamental solutions for the Klein-Gordon equation (\ref{Eqn_clKG}). The time flow of the Killing field determines a strongly continuous one-parameter unitary group $e^{itH}$ on $\mathcal{H}_e$ with a self-adjoint Hamiltonian $H$. If $H$ is invertible and the domain of $|H|^{-1}$ contains $E(f)$ for every test-function $f$ (this is the case e.g.\ when $m^2+\xi R>0$ everywhere) we can define quasi-free $\beta$-KMS states for all $\beta\in(0,\infty]$, where $\beta=\infty$ denotes the ground state. These are determined by the two-point distributions (acting on test-densities)
\begin{align}
\omega^{(\beta)}_2(f_1\sqrt{|\det g|},f_2\sqrt{|\det g|})&=
2\langle E(\overline{f_1}), H^{-1}\left(e^{\beta H}-1\right)^{-1}E(f_2)\rangle_e \,,\label{Eqn_KMS}\\
\omega^{(\infty)}_2(f_1\sqrt{|\det g|},f_2\sqrt{|\det g|})&=2\langle E(\overline{f_1}), P_-|H|^{-1}E(f_2)\rangle_e \,, \label{Eqn_ground}
\end{align}
where the inner products are taken in $\mathcal{H}_e$ and $P_-$ is the spectral projection for $H$ corresponding to the interval $(-\infty,0)$.

It is known that all ground and KMS states are Hadamard states \cite{Sahlmann+2000}. For any other state $\omega$ with a two-point distribution $\omega_2$, the difference $\omega_2-\omega_2^{(\infty)}$ is a symmetric and real-valued distribution on $M\times M$ which is a solution to the Klein-Gordon equation in each argument. The state $\omega$ is Hadamard if and only if the difference is smooth. Moreover, if the state is stationary, then $\omega_2-\omega_2^{(\infty)}$ is of positive type (cf. \cite{KS2017} Proposition 2.2).

Let us now turn to ground states on Einstein's static universe. For any fixed $a>0$, Einstein's static universe (\ref{Eqn_Estatic}) is an ultrastatic spacetime, whose symmetry group $\mathbb{R}\times \mathrm{SO}(4)$ consists of time-translations and rotations of the sphere $\mathbb{S}^3$. (These are all isometric diffeomorphisms that preserve the orientation and time orientation.)

It will be convenient to use the geodesic distance $\chi:\mathbb{S}^3\times\mathbb{S}^3\to\mathbb{R}$ between points on the unit sphere (using the metric $h$). Because $M$ is ultrastatic, its geodesic equation decouples into the geodesic equation for the time coordinate and the geodesic equation on $\mathbb{S}^3$ in the metric $a^2h_{ab}$. For the time coordinate, the geodesic distance is simply given by the Euclidean formula $|t-t'|$ for $t,t'\in\mathbb{R}$, whereas the geodesic distance on $\mathbb{S}^3$ in the metric $a^2h_{ab}$ is given by $a\chi$. It follows that Synge's world function, i.e.~half of the squared geodesic distance on $M$, can expressed  as
\begin{equation}\label{Eqn_Synge}
\sigma((t,x),(t',x'))=-\frac12(t-t')^2+\frac12a^2\chi(x,x')^2\,,
\end{equation}
where $x,x'\in\mathbb{S}^3$. This is a generalised form of Pythagoras' theorem. Note that $\chi$ and $\sigma$ are continuous functions, which are smooth as long as their arguments are not antipodal.

Again because $M$ is ultrastatic all time components of the Riemann curvature vanish, so we are essentially working with the Riemann curvature of the round sphere $(\mathbb{S}^3,a^2h)$. Because the Weyl tensor of every three-dimensional manifold vanishes and because the symmetries imply that $R_{ab}$ must be a multiple of $h_{ab}$ one can easily derive that
\begin{align}
R_{abcd}&=2a^2h_{c[a}h_{b]d}\,,\label{eqn:RiemannCurvature}\\
R_{ab}&=2h_{ab}\,,\label{eqn:RicciCurvature}\\
R&=\frac{6}{a^2}\,,\label{eqn:RicciScalar}
\end{align}
where the anti-symmetrisation is idempotent. The Einstein tensor therefore takes a perfect fluid form
\begin{align}\label{Eqn_G}
G_{ab}&=R_{ab}-\frac12Rg_{ab}=\frac{3}{a^2}(\nabla_at)(\nabla_bt)-h_{ab}\,.
\end{align}
The Klein-Gordon equation (\ref{Eqn_clKG}) simplifies to
\begin{align}\label{Eqn_KGinE}
(-\Box+m^2+\xi R)\varphi&=(\partial_t^2-a^{-2}\Delta_h+m^2+\xi R)\varphi=0\,,
\end{align}
where $\Delta_h$ is the Laplace-Beltrami operator on the unit sphere $\mathbb{S}^3$.

The two-point distribution $\omega_2^{(\infty)}$ of the ground state, given in Equation (\ref{Eqn_ground}) can be expressed in terms of functions of the operator
\begin{align}\label{eqn_defA}
A&:=(-\Delta_h+m^2a^2+6\xi)
\end{align}
on $L^2(\mathbb{S}^3,a^2h_{ab})$, which is essentially self-adjoint on $C_0^{\infty}(\mathbb{S}^3)$. The Hilbert space can be identified with $L^2(\mathbb{S}^3,h)$ for the unit sphere, which we will simply denote by $L^2(\mathbb{S}^3)$. The identification uses the unitary transformation $U$ defined by $Uf:=a^{\frac32}f$. The two-point distribution of the ground state is then given by
\begin{align}
\omega^{(\infty)}_2((t,\bar{f}_1a^3\sqrt{\det h}),(t',f_2a^3\sqrt{\det h}))&=
\frac12\left\langle f_1,(a^{-2}A)^{-\frac12}e^{-i(t-t')\sqrt{a^{-2}A}}f_2\right\rangle_{L^2(\mathbb{S}^3,a^2h_{ab})} \notag\\
&=\frac{a^4}{2}\left\langle f_1,A^{-\frac12}e^{-i\frac{t-t'}{a}\sqrt{A}}f_2\right\rangle_{L^2(\mathbb{S}^3)}\,, \label{Eqn_ground2}
\end{align}
where $f_1,f_2 \in C_0^{\infty}(\mathbb{S}^3)$. The self-adjoint operator $A$, viewed as an operator on $L^2(\mathbb{S}^3)$, has the same eigenspaces as $-\Delta_h$ and Appendix \ref{App_Harmonics} reviews how these eigenspaces may be determined explicitly. We will denote these eigenspaces by $L_n^{(3)}$ with $n\ge0$ and we let $E_n^{(3)}$ denote the orthogonal projection in $L^2(\mathbb{S}^3)$ onto $L_n^{(3)}$. The operator $A$ has corresponding eigenvalues $l_n^2$, where
\begin{align}
l_n&:=\sqrt{n(n+2)+m^2a^2+6\xi}\,,\label{eqn:defln}
\end{align}
so we may write $A=\sum_{n=0}^{\infty}l_n^2E_n^{(3)}$. By the functional calculus we have for all $t,t'\in\mathbb{R}$
\begin{align}
\omega^{(\infty)}_2((t,\bar{f}_1a^3\sqrt{\det h}),(t',f_2a^3\sqrt{\det h}))&=\sum_{n=0}^{\infty}
\frac{a^4}{2}\frac{1}{l_n}e^{-i\frac{t-t'}{a}l_n}\langle f_1,E_n^{(3)}f_2\rangle_{L^2(\mathbb{S}^3)}\,, \label{Eqn_ground3}
\end{align}
where $f_1,f_2 \in C_0^{\infty}(\mathbb{S}^3)$. At this point we can substitute the integral kernel $E_n^{(3)}(x,x')$ of $E_n^{(3)}$, given in terms of Gegenbauer polynomials $C_n^{(1)}$ in Equation (\ref{eqn:projectionkernel}), and write
\begin{align}
\omega^{(\infty)}_2((t,x),(t',x'))&=\lim_{\epsilon\to0^+}\sum_{n=0}^{\infty}\frac{1}{2a^2}\frac{1}{l_n}e^{-i\frac{t-t'-i\epsilon}{a}l_n}E_n^{(3)}(x,x')\label{Eqn_ground4}\\
&=\lim_{\epsilon\to0^+}\sum_{n=0}^{\infty}\frac{1}{4\pi^2a^2}\frac{n+1}{l_n}e^{-i\frac{t-t'-i\epsilon}{a}l_n}C_n^{(1)}(\cos(\chi(x,x')))\,,\notag
\end{align}
where we introduced $\epsilon>0$ to regularise the sum.


In the special case when $m^2a^2+6\xi=1$, e.g.~when $m=0$ and $\xi=\xi_c=\frac16$, Equation (\ref{Eqn_ground4}) simplifies, because $l_n=n+1$. We may then use Equation (\ref{eqn:generating}) and find (cf.~\cite{Birrel+} Eqn.(5.32))
\begin{align}
\omega^{(\infty)}_2((t,x),(t',x'))&=\lim_{\epsilon\to0^+}\sum_{n=0}^{\infty}
\frac{1}{2a^2}\frac{1}{n+1}e^{-i\frac{t-t'-i\epsilon}{a}(n+1)}E_n^{(3)}(x,x')\notag\\
&=\lim_{\epsilon\to0^+}\frac{1}{8\pi^2a^2}
\frac{1}{\cos\left(\frac{t-t'}{a}-i\epsilon\right)-\cos(\chi(x,x'))}\,.\label{Eqn_ground5}
\end{align}

For the general case, without assuming $m^2a^2+6\xi=1$, the ground states are more complicated and we will not study them in full detail. However, it will be useful to consider the restriction of $\omega^{(\infty)}_2$ to $x=x'$, which is 
\begin{align}
\omega^{(\infty)}_2((t,x),(t',x))&=\lim_{\epsilon\to0^+}\sum_{n=0}^{\infty}\frac{1}{4\pi^2a^2}
\frac{(n+1)^2}{l_n}e^{-i\frac{t-t'-i\epsilon}{a}l_n}\,.\label{eqn:restrictedgroundstate}
\end{align}
To analyse this series expansion we can write
$e^{-i\frac{t-t'-i\epsilon}{a}l_n}=e^{-i\frac{t-t'-i\epsilon}{a}(l_n-n-1)}e^{-i\frac{t-t'-i\epsilon}{a}(n+1)}$
and then expand $\frac{(n+1)^2}{l_n}e^{-i\frac{t-t'-i\epsilon}{a}(l_n-n-1)}$ in powers of $(n+1)^{-1}$. Sums of the form
$\sum_{n=0}^{\infty}(n+1)^{-k}z^{n+1}$ with $k\in\mathbb{Z}$ and $z=e^{-i\frac{t-t'-i\epsilon}{a}}$ can be computed using integrals or derivatives of a geometric series, because $|z|<1$. In the limit $\epsilon\to0^+$ we then see that sufficiently large powers $k$ give rise to continuous contributions, whereas for small or negative $k$ we find distributions that are not given by continuous functions.

The tedious details of this analysis are given in Appendix \ref{app:ground} and we only record the results here. In terms of the parameter
\begin{align}
c&:=m^2a^2+6\xi-1\label{eqn:defc}
\end{align}
and
\begin{align}
T_{\epsilon}&:=\frac{t-t'-i\epsilon}{a}\notag
\end{align}
with $\epsilon>0$ we have
\begin{align}
\omega^{(\infty)}_2((t,x),(t',x))&=\lim_{\epsilon\to0^+}\frac{1}{4\pi^2a^2}\left(\frac{-1}{T_{\epsilon}^2}+\frac{c}{4}
\log(-T_{\epsilon}^2)\right)+\frac{1}{4\pi^2a^2}X_1+O(t-t')\label{eqn:restrictedgroundstate2}
\end{align}
where
\begin{align}
X_1&=-\frac{1+6c}{12}+
\sum_{n=0}^{\infty}\frac{(n+1)^2}{l_n}-(n+1)+\frac{c}{2(n+1)}\,.\label{eqn:X1}
\end{align}

Using similar methods we can analyse
\begin{align}
\partial_t\partial_{t'}\omega^{(\infty)}_2((t,x),(t',x))&=\lim_{\epsilon\to0^+}\sum_{n=0}^{\infty}\frac{1}{4\pi^2a^4}
(n+1)^2l_ne^{-iT_{\epsilon}l_n}\label{eqn:00}
\end{align}
which leads to
\begin{align}
\partial_t\partial_{t'}\omega^{(\infty)}_2((t,x),(t',x))&=
\lim_{\epsilon\to0^+}\frac{1}{4\pi^2a^4}\left(\frac{6}{T_{\epsilon}^4}+\frac{c}{2T_{\epsilon}^2}
+\frac{c^2}{16}\log(-T_{\epsilon}^2)\right)+\frac{1}{4\pi^2a^4}X_2+O(t-t')\label{eqn:restrictedgroundstate3}
\end{align}
where
\begin{align}
X_2&=\frac{1+10c-15c^2}{120}+
\sum_{n=0}^{\infty}(n+1)^2l_n-(n+1)^3-\frac{c}{2}(n+1)+\frac{c^2}{8(n+1)}\,. \label{eqn:X2}
\end{align}
Again the details of this analysis are given in Appendix \ref{app:ground}.

\section{Symmetric quasi-free states}\label{Sec_SymStates}

We will call a two-point distribution $\omega_2$ of a quantum state symmetric when it is invariant under all the spacetime symmetries, i.e.\ when
\begin{align}\label{Eqn_DefSym}
\omega_2(R(t,x),R(t',x'))&=\omega_2((t,x),(t',x'))
\end{align}
whenever $R:M\to M$ is a symmetry. We will now characterise all such symmetric two-point distributions without assuming the Hadamard property.

\begin{theorem}\label{Thm_sym}
Consider an Einstein static universe of radius $a>0$ and a real free scalar quantum field of mass $m\ge0$ and scalar curvature coupling $\xi\in\mathbb{R}$ such that $m^2+\xi R>0$. For every symmetric two-point distribution $\omega_2$ in the sense of (\ref{Eqn_DefSym}) we have:
\begin{align}\label{Eqn_SymState}
\omega_2((t,x),(t',x'))&=\omega^{\infty}_2((t,x),(t',x')) + \sum_{n=0}^{\infty}\frac{a_n}{n+1}\cos((t-t')a^{-1}l_n)C_n^{(1)}(\cos(\chi(x,x'))) \,,
\end{align}
where $C_n^{(1)}$ is a Gegenbauer polynomial (\ref{eqn:Rodrigues}), $l_n$ is given in (\ref{eqn:defln}) and $a_n\ge0$.
\end{theorem}
\begin{proof*}
The two-point distribution $\omega^{(\infty)}_2$ of the ground state is itself symmetric. To prove (\ref{Eqn_SymState}) for a general symmetric two-point distribution $\omega_2$, we first note that $\omega_2$ is in particular stationary, so by Proposition 2.2 of \cite{KS2017} the distribution
\begin{align}\label{Eqn_Defw2}
w_2&:=\omega_2-\omega^{(\infty)}_2
\end{align}
on $M\times M$ is a real-valued bisolution to the Klein-Gordon equation which is of positive type, invariant under the exchange of its two arguments and under the spacetime symmetries.

Because $w_2$ is a bisolution to the Klein-Gordon equation (\ref{Eqn_clKG}) its singularities are light-like, so we can restrict $w_2$ and its derivatives by setting $t=0$ and/or $t'=0$ by standard microlocal arguments \cite{Hormander}. In particular, $w_2((0,x),(0,x'))$ is of positive type, so it defines a positive quadratic form on $C_0^{\infty}(\mathbb{S}^3)$ in the Hilbert space
$L^2(\mathbb{S}^3)$. This quadratic form has a self-adjoint Friedrichs extension $W$, which is again positive, so we have $W\ge0$ and
\begin{align}
w_2((0,f_1a^3\sqrt{\det h}),(0,f_2a^3\sqrt{\det h}))&=\langle \bar{f}_1, Wf_2\rangle_{L^2(\mathbb{S}^3)}
\end{align}
for all $f_1,f_2\in C_0^{\infty}(\mathbb{S}^3)$. Note that $w_2((0,x),(0,x'))$ and hence also $W$ is invariant under rotations of $\mathbb{S}^3$.

For every $\eta_1,\eta_2\in C_0^{\infty}(\mathbb{R})$ the distribution $w_2((\eta_1,x),(\eta_2,x'))$ is a smooth function of $(x,x')$, using again that $w_2$ is a bisolution to the Klein-Gordon equation \cite{Hormander}. The invariance under exchange of arguments and under rotations of $\mathbb{S}^3$ then imply
\begin{align}
w_2((\eta_1,x),(\eta_2,x'))&=w_2((\eta_2,x'),(\eta_1,x))=w_2((\eta_2,x),(\eta_1,x')) \,,
\end{align}
i.e.\ $w_2((t,x),(t',x'))$ is invariant under exchanging $t$ and $t'$. Using the stationarity we then see from
$w_2((t-t',x),(0,x'))=w_2((t,x),(t',x'))=w_2((t',x),(t,x'))=w_2((t'-t,x),(0,x'))$ that
\begin{align}\label{Eqn_w10}
\partial_tw_2((t,x),(t',x'))|_{t=t'=0}&=\partial_{t'}w_2((t,x),(t',x'))|_{t=t'=0}=0 \,.
\end{align}
Furthermore,
\begin{align}\label{Eqn_w11}
\partial_{t'}\partial_tw_2((t,x),(t',x'))|_{t=t'=0}&=-\partial_t^2w_2((t,x),(t',x'))|_{t=t'=0}\notag\\
&=a^{-2}(-\Delta_h+m^2a^2+6\xi)w_2((t,x),(t',x'))|_{t=t'=0} \\
&=-\partial_{t'}^2w_2((t,x),(t',x'))|_{t=t'=0}\notag\\
&=a^{-2}(-\Delta'_h+m^2a^2+6\xi)w_2((t,x),(t',x'))|_{t=t'=0} \,,\notag
\end{align}
where we used the Klein-Gordon operator (\ref{Eqn_KGinE}), $R=6a^{-2}$ and $\Delta'_h$ is the Laplace-Beltrami operator acting on the variables $x'$. This means that the operator $W$ commutes with the operator $A$ of (\ref{eqn_defA}) on $\mathbb{S}^3$.

In particular, $W$ preserves the eigenspaces $L_n^{(3)}$ of $-\Delta_h$. Because $-\Delta_h$ commutes with the action of the group $SO(4,\mathbb{R})$, there is a representation of this group on each eigenspace $L_n^{(3)}$. The fact that these representations are irreducible is reviewed in Appendix \ref{App_Harmonics}. The operator $W$ restricted to any $L_n^{(3)}$ can be diagonalised. However, because $W$ also commutes with the action of $SO(4,\mathbb{R})$, its eigenspaces in $L_n^{(3)}$ must be invariant under $SO(4,\mathbb{R})$. By the irreducibility and the positivity of $W$ this means that $W$ acts as a non-negative multiple of the identity on each $L_n^{(3)}$. It follows that
\begin{align}\label{Eqn_w00}
w_2((0,x),(0,x'))&=\sum_{n=0}^{\infty}\frac{2\pi^2a_n}{(n+1)^2}E_n^{(3)}(x,x')
=\sum_{n=0}^{\infty}\frac{a_n}{n+1}C_n^{(1)}(\cos(\chi(x,x')))
\end{align}
with $a_n\ge0$, where we used Equation (\ref{eqn:projectionkernel}) with $p=3$ to express the integral kernel $E_n^{(3)}(x,x')$ of the orthogonal projection $E_n^{(3)}$ onto $L_n^{(3)}$ in terms of the Gegenbauer polynomial $C_n^{(1)}$.


Because $w_2$ is a bisolution to the Klein-Gordon equation, it is uniquely determined by the initial data in (\ref{Eqn_w00}, \ref{Eqn_w10}, \ref{Eqn_w11}), i.e.\ it is uniquely determined by the operator $W$. Because $C_n^{(1)}$ projects onto an eigenspace of $-\Delta_h$ with eigenvalue $n(n+2)$ one may easily verify that $\omega_2$ must take the form stated in Equation (\ref{Eqn_SymState}). Conversely, any two-point distribution of this form has all the desired properties.
\end{proof*}

A two-point distribution $\omega_2$ is called Hadamard when the difference $w_2=\omega_2-\omega_2^{(\infty)}$ is smooth. For symmetric two-point distributions this means that the series in Equation (\ref{Eqn_SymState}) must converge to a smooth function. This is the case when $\sum_{n=0}^{\infty}a_nl_n^k$ converges for each $k\in\mathbb{N}$. Because $l_n\sim n$ for large $n$ this means that the coefficients $a_n$ must fall off faster than any polynomial in $n$.

It is worth noting that the two-point distributions $\omega_2^{(\beta)}$ of the $\beta$-KMS states, given in (\ref{Eqn_KMS}), can also be expressed in terms of functions of the operator $a^{-2}A$ on $L^2(\mathbb{S}^3,a^2h_{ab})$. Using similar arguments as for the ground state, cf.\ Equation (\ref{Eqn_ground2}), we find
\begin{align}
\omega^{(\beta)}_2((t,\bar{f}_1a^3\sqrt{\det h}),(t',f_2a^3\sqrt{\det h}))&=
\frac12\left\langle f_1, (a^{-2}A)^{-\frac12}\frac{\cos\left((t-t'+i\frac{\beta}{2})\sqrt{a^{-2}A}\right)}
{\sinh\left(\frac{\beta}{2}\sqrt{a^{-2}A}\right)}f_2\right\rangle_{L^2(\mathbb{S}^3,a^2h_{ab})} \notag\\
&=\frac{a^4}{2}\left\langle f_1, A^{-\frac12}\frac{\cos\left((\frac{t-t'}{a}+i\frac{\beta}{2a})\sqrt{A}\right)}
{\sinh\left(\frac{\beta}{2a}\sqrt{A}\right)}f_2\right\rangle_{L^2(\mathbb{S}^3)} \,, \label{Eqn_KMS2}
\end{align}
where $f_1,f_2 \in C_0^{\infty}(\mathbb{S}^3)$.
%
Comparing with Theorem \ref{Thm_sym} and using Equation (\ref{eqn:projectionkernel}) in the appendix we then see that the
$\omega^{(\beta)}_2$ are determined by the values
\begin{align}\label{eqn_thermalAn}
a_n^{(\beta)}=\frac{(n+1)^2}{2\pi^2 a^3}\ \frac{a}{l_n}\ \frac{e^{-\beta\frac{l_n}{a}}}{1-e^{-\beta\frac{l_n}{a}}}\,.
\end{align}
This may be seen by setting $t_0=t_1=0$ in Equations (\ref{Eqn_ground2}, \ref{Eqn_KMS2}) and writing the function of $A$ in terms of the eigenspaces and eigenvalues of $-\Delta_h$.

\section{The stress-energy-momentum tensor}\label{Sec_Stress}

To renormalise the quantum stress tensor we will use the Hadamard prescription. For this purpose we will first review the properties of the Hadamard parametrix. Due to the large amount of symmetry of the Einstein static universe, this series takes a relatively simple form in this case.

The general form of the Hadamard series is
\begin{align}\label{Eqn_Hadseries1}
H:=\frac{u_0}{8\pi^2\sigma_+}+\frac{1}{8\pi^2}\sum_{k=1}^{\infty}\frac{(-2)^{-k}u_k}{(k-1)!}\sigma_+^{k-1}\log(\sigma_+)\,,
\end{align}
where we use the notation $F(\sigma_+):=\lim_{\epsilon\to 0^+}F(\sigma_{\epsilon})$ for suitable functions $F$ and taking a distributional limit with
\begin{align}\label{Eqn_sigma_eps}
\sigma_{\epsilon}((t,x),(t',x'))&:=-\frac12(t-t'-i\epsilon)^2+\frac{a^2}{2}\chi(x,x')^2
\end{align}
for any $\epsilon>0$. The Hadamard coefficients $u_k$ are smooth functions on a neighbourhood of the diagonal
\begin{align}
D&:=\{(x,x)|\ x\in M\}\subset M\times M\label{eqn:defD}
\end{align}
satisfying
\begin{align}\label{Eqn_transport}
(-\Box+m^2+\xi R)u_{k-1}&=-\nabla^{\mu}\sigma\nabla_{\mu}u_k-\left(k+\frac12\Box\sigma-2\right)u_k
\end{align}
with $u_{-1}\equiv 0$ and $u_0|_D\equiv 1$. One may show that the Hadamard coefficients are symmetric under exchange of their arguments \cite{Moretti2000}. The importance of the Hadamard series is that its singularities near the diagonal are fully determined by the local geometry and the parameters $m$ and $\xi$ and they coincide exactly with the singularities of all Hadamard two-point distributions, including ground and KMS states.

\begin{remark}\label{rem:Hadscale}
We should point out that $\sigma$ is a dimensionful quantity, so to make the logarithm in (\ref{Eqn_Hadseries1}) well-defined we should really use $\log(\sigma_+/\ell^2)=\log(\sigma_+)-\log(\ell^2)$ for some length $\ell$. The choice of this length scale introduces some ambiguity in the Hadamard regularisation prescription. Instead of explicitly including this length scale here, we note that the resulting ambiguity is also included in the renormalisation freedom of the stress tensor in Equation (\ref{Eqn_Tren}) below, cf.~Thm.~2.6.1 of \cite{Hack2016} and also \cite{Moretti2003}.
\end{remark}

In the Einstein static universe the Hadamard coefficients are given in terms of the parameter $c$ of (\ref{eqn:defc}) by
\begin{align}
u_k((t,x),(t',x'))&=\frac{(-c)^k}{a^{2k}k!}\frac{\chi(x,x')}{\sin(\chi(x,x'))}
\end{align}
for all $k\in\mathbb{N}_0$. These Hadamard coefficients are defined on the set $N=\{((t,x),(t',x'))\in M\times M\mid \chi(x,x')<\pi\}$ where $x$ and $x'$ are not antipodal.
%
The Hadamard series (\ref{Eqn_Hadseries1}) then simplifies to
\begin{align}\label{Eqn_Hadseries2}
H&=\frac{u_0}{8\pi^2\sigma_+}+\frac{u_0c}{16\pi^2a^2}\log(\sigma_+)
\sum_{k=0}^{\infty}\frac{1}{k!(k+1)!}\left(\frac{c}{2a^2}\sigma_+\right)^k \,,
\end{align}
which converges (in the sense of distributions). In fact, we may distinguish three cases. When $c=0$, e.g.\ in the conformally invariant case $m=0$, $\xi=\xi_c$, the Hadamard series simplifies to
\begin{align}\label{Eqn_Hadcase0}
H&=\frac{u_0}{8\pi^2\sigma_+} \,.
\end{align}
%
When $c>0$, then 
\begin{align}\label{Eqn_Hadcase>}
H&=\frac{u_0}{8\pi^2\sigma_+}+\frac{u_0}{16\pi^2a\sigma_+}\sqrt{2c\sigma_+}\log(\sigma_+)
I_1\left(\frac{\sqrt{2c\sigma_+}}{a}\right) \,,
\end{align}
where $I_1$ is a modified Bessel function (cf. \cite{Gradshteyn+} (8.445)).
%
%
When $c<0$, then
\begin{align}\label{Eqn_Hadcase<}
H&=\frac{u_0}{8\pi^2\sigma_+}
-\frac{u_0}{16\pi^2a\sigma_+}\sqrt{-2c\sigma_+}\log(\sigma_+)J_1\left(\frac{\sqrt{-2c\sigma_+}}{a}\right) \,,
\end{align}
where $J_1$ is a Bessel function of the first kind (cf. \cite{Gradshteyn+} Equation (8.402)).
%

In general, the Hadamard series is not a bisolution of the Klein-Gordon equation. Instead we have
\begin{align}
(-\Box+m^2+\xi R)H&=
\frac{-u_0c^2}{16\pi^2a^4}\sum_{k=0}^{\infty}\frac{(2k+3)}{(k+1)!(k+2)!}\left(\frac{c\sigma}{2a^2}\right)^k \,,
\end{align}
which vanishes only if $c=0$. On the diagonal $D$ we have in particular
\begin{align}\label{Eqn_Hadsource}
(-\Box+m^2+\xi R)H|_D&=\frac{-3c^2}{32\pi^2a^4}=\frac{-3}{32\pi^2}(m^2+(\xi-\xi_c)R)^2\,.
\end{align}

In general curved spacetimes one may use the Hadamard parametrix to renormalise the stress tensor as follows. Splitting the points in the classical definition (\ref{Eqn_T}) we may write
\begin{align}\label{Eqn_Tsplit}
T_{ab}(x)&=\lim_{x'\to x}D^{\mathrm{split}}_{ab}\varphi(x)\varphi(x')
+\xi \left(G_{ab}(x)-\nabla_a\nabla_b+g_{ab}(x)\Box\right)\lim_{x'\to x}\varphi(x)\varphi(x') \,,
\end{align}
where $D^{\mathrm{split}}_{ab}$ is the differential operator defined in a neighbourhood of the diagonal $D$ in $M\times M$ by
\begin{align}\label{Eqn_Dsplit}
D^{\mathrm{split}}_{ab}:=&\frac12\delta^{b'}_b\nabla_a\nabla'_{b'}+\frac12\delta^{a'}_a\nabla'_{a'}\nabla_b
-\frac12g_{ab}(x)\delta^{c'}_c\nabla^c\nabla'_{c'}-\frac12g_{ab}m^2 \,.
\end{align}
Here $\delta^{a'}_a$ denotes the parallel transport of tangent vectors along the geodesic from $x$ to $x'$, which is well-defined when $x'$ is close enough to $x$. In Equation (\ref{Eqn_Tsplit}) we now replace the classical field $\varphi$ by the quantum field $\phi$, symmetrise and regularise by subtracting $H_+(x,x'):=\frac12(H(x,x')+H(x',x))$ (as a multiple of the identity operator):
\begin{align}\label{Eqn_Treg}
T^{\mathrm{reg}}_{ab}(x):=&\lim_{x'\to x}D^{\mathrm{split}}_{ab}\left(\frac12\phi(x)\phi(x')+\frac12\phi(x')\phi(x)-H_+(x,x')\right)\notag\\
&+\xi \left(G_{ab}(x)-\nabla_a\nabla_b+g_{ab}(x)\Box\right)\lim_{x'\to x}\left(\frac12\phi(x)\phi(x')+\frac12\phi(x')\phi(x)-H_+(x,x')\right) \,. 
\end{align}
Here the limit is to be understood in a suitable topology, which we will not describe in detail. For us it is sufficient to know that the expectation value on the right-hand side converges in any state for which $\omega_{2+}-H_+$ is twice differentiable. This certainly includes all Hadamard states.

Although $\omega(T^{\mathrm{reg}}_{ab})$ is a smooth symmetric tensor field for every Hadamard state $\omega$, it is in general not divergence free in the sense that \cite{Moretti2003}
\begin{align}
\omega(\nabla^aT^{\mathrm{reg}}_{ab})&=\frac{-1}{16\pi^2}\nabla_b(u_2|_D)\,.\label{eqn:divu2}
\end{align}
Furthermore, we might replace $H$ by some other local and covariant expression with the same singularity structure to regularise the stress tensor. For this reason we introduce the renormalised stress tensor\footnote{The $\Lambda$-dependent renormalisation ambiguities are often omitted. In the semi-classical Einstein equation they can be absorbed in a renormalised cosmological constant and Newton's constant, as we will see in Section \ref{sec_SSE}. One may also omit the term $\alpha_5K_{ab}$, because $K_{ab}=4J_{ab}-I_{ab}$, cf.~\cite{Hack2016}. However, in our case of interest we will see even stronger reductions due to the symmetry.}
\begin{align}\label{Eqn_Tren}
T^{\mathrm{ren}}_{ab}:=&T^{\mathrm{reg}}_{ab}+g_{ab}\frac{u_2|_D}{16\pi^2}\notag\\
&+\alpha_1m^4g_{ab}+\alpha_2m^2G_{ab}+\alpha_3I_{ab}+\alpha_4J_{ab}+\alpha_5K_{ab}\\
&+\beta_1\Lambda^2g_{ab}+\beta_2\Lambda G_{ab}+\beta_3\Lambda m^2g_{ab}\,.\notag
\end{align}
The second term on the first line of (\ref{Eqn_Tren}) ensures that the first line is divergence free. Explicitly one may compute (cf.\ \cite{Decanini+2008}
Eqn.(109) and note that the coefficients $v_k$ in that reference are given by $v_k=(-2)^{-k-1}u_{k+1}$)
\begin{align}
u_2|_D&=\frac{c^2}{2a^4}+\frac16\left(\frac15-\xi\right)\Box R-\frac{1}{180}R_{ab}R^{ab}+\frac{1}{180}R_{abcd}R^{abcd}\,.
\end{align}
The terms on the second and third lines of (\ref{Eqn_Tren}) are all local, divergence free, symmetric tensors of type $(0,2)$ with the same scaling behaviour as the stress tensor \cite{Hollands+2005,Hack2016}. The coefficients $\alpha_i$ and $\beta_i$ are analytic functions of the dimensionless parameter $\xi$ and the terms $I_{ab}$, $J_{ab}$ and $K_{ab}$ can be written as variational derivatives w.r.t.\ $g^{ab}$ of the Lagrangian densities $R^2\sqrt{\mathrm{det} g}$,  $R_{cd}R^{cd}\sqrt{\mathrm{det} g}$ and $R_{cdef}R^{cdef}\sqrt{\mathrm{det} g}$, respectively. Explicitly we have (cf.\ \cite{Hack2016} Eqns.(2.28-29) and note that we have corrected some sign errors):
\begin{align}
I_{ab}&=2RR_{ab}-2\nabla_a\nabla_bR-\frac12g_{ab}R^2+2g_{ab}\Box R \label{Eqn_I}\\
J_{ab}&=\Box R_{ab}-\nabla_a\nabla_bR+2R_{acbd}R^{cd}-\frac12g_{ab}R_{cd}R^{cd}+\frac12g_{ab}\Box R \label{Eqn_J}\\
K_{ab}&=4\Box R_{ab}-2\nabla_a\nabla_bR+2R_{acde}R_b^{\phantom{b}cde}+4R_{acbd}R^{cd}-4R_{ac}R^c_{\phantom{c}b}
-\frac12g_{ab}R_{cdef}R^{cdef} \label{Eqn_K}\,.
\end{align}

In an Einstein static universe the general formula (\ref{Eqn_Tren}) simplifies considerably due to the symmetries of the curvature tensors. Equations (\ref{eqn:RiemannCurvature}, \ref{eqn:RicciCurvature}, \ref{eqn:RicciScalar}) imply that
\begin{align}\label{Eqn_u2simple}
u_2|_D&=\frac{c^2}{2a^4}
\end{align}
and
\begin{align}\label{Eqn_IJKsimple}
I_{ab}=3J_{ab}=3K_{ab}&=2RR_{ab}-\frac12g_{ab}R^2=2RG_{ab}+\frac12g_{ab}R^2 \,.
\end{align}
Because $u_2|_D$ is constant we see from (\ref{eqn:divu2}) that $T^{\mathrm{reg}}_{ab}$ is already conserved and we can parameterise the renormalisation freedom as
\begin{align}\label{Eqn_TrenE}
T^{\mathrm{ren}}_{ab}=&T^{\mathrm{reg}}_{ab}+\frac{c^2}{32\pi^2a^4}g_{ab}+c_1g_{ab}+c_2R_{ab} \,, 
\end{align}
where
\begin{align}
c_1(\xi,m,\Lambda,a)=&\alpha_1m^4+\beta_1\Lambda^2+\beta_3\Lambda m^2 -\frac12R(\alpha_2m^2+\beta_2\Lambda)
-\frac16R^2\left(3\alpha_3+\alpha_4+\alpha_5\right) \,, \label{Eqn_c1}\\
c_2(\xi,m,\Lambda,a)=&\alpha_2m^2+\beta_2\Lambda+\frac23 R\left(3\alpha_3+\alpha_4+\alpha_5\right) \,. \label{Eqn_c2}
\end{align}

For any state $\omega$ for which $\omega_{2+}-H_+$ is symmetric and $C^2$ we may use the fact that $(\omega_{2+}-H_+)|_D$ is constant to simplify the 
expectation value of (\ref{Eqn_Treg}) to 
\begin{align}\label{Eqn_TregExp1}
\omega(T^{\mathrm{reg}}_{ab})=&\lim_{x'\to x}D^{\mathrm{split}}_{ab}(\omega_{2+}-H_+)(x,x')
+\xi(\omega_{2+}-H_+)|_DG_{ab} \ .
\end{align}
The right-hand side of (\ref{Eqn_TregExp1}) is a continuous tensor field which is symmetric. This means in particular that
for states $\omega$ with a symmetric two-point distribution,
\begin{align}\label{Eqn_TregExp2}
\omega(T^{\mathrm{reg}}_{ab})&=E^{\mathrm{reg}}(\omega)\nabla_at\nabla_bt-P^{\mathrm{reg}}(\omega)a^2h_{ab}
\end{align}
has a perfect fluid form with a constant regularised energy $E^{\mathrm{reg}}(\omega)$ and pressure $P^{\mathrm{reg}}(\omega)$.
In view of (\ref{Eqn_TrenE}) the renormalised stress tensor also has a perfect fluid form
\begin{align}\label{Eqn_Tren2}
\omega(T^{\mathrm{ren}}_{ab})&=E^{\mathrm{ren}}(\omega)\nabla_at\nabla_bt-P^{\mathrm{ren}}(\omega)a^2h_{ab}
\end{align}
where the renormalised energy and pressure are
\begin{align}
E^{\mathrm{ren}}(\omega)&=E^{\mathrm{reg}}(\omega)-\frac{c^2}{32\pi^2a^4}-c_1\,,\label{eqn:Eren}\\
P^{\mathrm{ren}}(\omega)&=P^{\mathrm{reg}}(\omega)-\frac{c^2}{32\pi^2a^4}-c_1-\frac13c_2R\,.\label{eqn:Pren}
\end{align}

At this point it is helpful to use the following
\begin{lemma}\label{Lem_swap}
Let $w$ be a $C^2$ function on a neighbourhood $U$ of the diagonal $D$ (see (\ref{eqn:defD})) and assume that $w$ is symmetric, i.e.~$w(R(t,x),R(t',x'))=w((t,x),(t',x'))$ whenever $R:M\to M$ is a symmetry and both $((t,x),(t',x'))\in U$ and $(R(t,x),R(t',x'))\in U$. Then
\begin{align}
\nabla_b\nabla'_aw((t,x),(t',x'))|_D&=-\nabla_a\nabla_bw((t,x),(t',x'))|_D\,.
\end{align}
\end{lemma}
\begin{proof*}
We will use local coordinates $\{x^i\}_{i=1}^3$ on $\mathbb{S}^3$ and the Killing time coordinate $t$. From the symmetry we then see that $w((t,x),(t',x))$ only depends on $t-t'$ and that $\partial_tw|_D$ is in fact a constant function on $M$. Furthermore, the components $\partial_iw(x,x')|_D$ define a dual vector field in $T^*M$ which annihilates the timelike Killing field. Due to the symmetry, this dual vector cannot pick out any preferred spatial direction, so we must have $\partial_iw(x,x')|_D\equiv 0$. Because all components of $\partial_{\nu}w(x,x')|_D$ are constant we have for any spacetime indices $\mu,\nu$,
\begin{align}
0&=\partial_{\mu}\left(\partial_{\nu}w(x,x')|_{x'=x}\right)=\partial_{\mu}\partial_{\nu}w(x,x')|_D+\partial'_{\mu}\partial_{\nu}w(x,x')|_D \,.
\end{align}
We may replace the coordinate derivatives with covariant derivatives, because the terms involving Christoffel symbols are multiplied either by $\partial_iw(x,x')|_D\equiv 0$ or by $\Gamma^0_{\mu\nu}\equiv 0$. The lemma then follows.
\end{proof*}

From (\ref{Eqn_TregExp2}, \ref{Eqn_TregExp1}) and Lemma \ref{Lem_swap} we find that
\begin{align}\label{Eqn_00}
E^{\mathrm{reg}}(\omega)&=\omega(T^{\mathrm{reg}}_{00})\notag\\
&=D^{\mathrm{split}}_{00}(\omega_{2+}-H_+)|_D
+\frac12\xi R(\omega_{2+}-H_+)|_D \notag\\
&=\left(\partial_t\partial_{t'}+\frac12(-\Box+m^2+\xi R)\right)(\omega_{2+}-H_+)|_D\notag\\
&=\partial_t\partial_{t'}(\omega_{2+}-H_+)|_D+\frac{3c^2}{64\pi^2a^4}\,.
\end{align}
where we used the fact that $\omega_{2+}$ is a bisolution of the Klein-Gordon equation and (\ref{Eqn_Hadsource}) in the last line.
Similarly,
\begin{align}\label{Eqn_trace}
E^{\mathrm{reg}}(\omega)+3P^{\mathrm{reg}}(\omega)&=-g^{ab}\omega(T^{\mathrm{reg}}_{ab})\notag\\
&=-g^{ab}D^{\mathrm{split}}_{ab}(\omega_{2+}-H_+)|_D
+\xi R(\omega_{2+}-H_+)|_D \notag\\
&=-g^{ab}\left(-\nabla_a\nabla_b+\frac12g_{ab}\Box-\frac12g_{ab}m^2\right)(\omega_{2+}-H_+)|_D
+\xi R(\omega_{2+}-H_+)|_D \notag\\
&=(-\Box+2m^2+\xi R)(\omega_{2+}-H_+)|_D\notag\\
&=m^2(\omega_{2+}-H_+)|_D +\frac{3c^2}{32\pi^2a^4}
\end{align}
and therefore
\begin{align}\label{Eqn_Preg}
P^{\mathrm{reg}}(\omega)
&=\frac13(m^2-\partial_t\partial_{t'})(\omega_{2+}-H_+)|_D+\frac{c^2}{64\pi^2a^4}\,.
\end{align}
For the renormalised energy and pressure we see from (\ref{eqn:Eren}) and (\ref{eqn:Pren}) that
\begin{align}
E^{\mathrm{ren}}(\omega)&=\partial_t\partial_{t'}(\omega_{2+}-H_+)|_D+\frac{c^2}{64\pi^2a^4}-c_1\,,\label{eqn:Eren2}\\
P^{\mathrm{ren}}(\omega)
&=\frac13(m^2-\partial_t\partial_{t'})(\omega_{2+}-H_+)|_D-\frac{c^2}{64\pi^2a^4}-c_1-\frac13c_2R\,. \label{eqn:Pren2}
\end{align}
Note that
\begin{align}\label{Eqn_trace2}
E^{\mathrm{ren}}(\omega)+3P^{\mathrm{ren}}(\omega)&=-g^{ab}\omega(T^{\mathrm{ren}}_{ab})
=m^2(\omega_{2+}-H_+)|_D-\frac{c^2}{32\pi^2a^4}-4c_1-c_2R\,,
\end{align}
which becomes independent of the choice of the symmetric two-point distribution when $m=0$.

Let us now investigate the regularised energy density for the ground state. For simplicity we first consider the special case where $c=0$. In this case the ground state is given explicitly in (\ref{Eqn_ground5}) and the Hadamard series consists of the single term (\ref{Eqn_Hadcase0}), so the difference is
\begin{align}
(\omega^{(\infty)}_2-H)((t,x),(t',x'))&=\frac{1}{8\pi^2a^2}\lim_{\epsilon\to0^+}
\frac{1}{\cos\left(\frac{t-t'-i\epsilon}{a}\right)-\cos(\chi(x,x'))}
-\frac{u_0(x,x')}{-\frac12\left(\frac{t-t'-i\epsilon}{a}\right)^2+\frac12\chi(x,x')^2}\,.\notag
\end{align}
One can show that this is smooth near the diagonal $(t,x)=(t',x')$. To compute the energy density we first restrict to $x=x'$, where we can expand the cosine function in a Taylor series. In terms of $T_{\epsilon}=\frac{t-t'-i\epsilon}{a}$ we find
\begin{align}
(\omega^{(\infty)}_2-H)((t,x),(t',x))&=\frac{1}{8\pi^2a^2}\lim_{\epsilon\to0^+}
\frac{1}{\cos(T_{\epsilon})-1}-\frac{1}{-\frac12T_{\epsilon}^2}\notag\\
&=\frac{1}{8\pi^2a^2}\lim_{\epsilon\to0^+}
\frac{1}{-\frac12T_{\epsilon}^2}\frac{1}{1-\frac{2}{4!}T_{\epsilon}^2+\frac{2}{6!}T_{\epsilon}^4
+O(T_{\epsilon}^6)}-\frac{1}{-\frac12T_{\epsilon}^2}\notag\\
&=\frac{1}{8\pi^2a^2}\lim_{\epsilon\to0^+}
\frac{1}{-\frac12T_{\epsilon}^2}\left(\frac{2}{4!}T_{\epsilon}^2+\frac{3T_{\epsilon}^4}{6!}+O(T_{\epsilon}^6)\right)\notag\\
&=\frac{-1}{48\pi^2a^2}-\frac{1}{960\pi^2a^4}(t-t')^2+O((t-t')^4)\,.\label{eqn:conformalcase}
\end{align}
Note that this result is automatically symmetric under exchange of $t$ and $t'$, so we can now compute
\begin{align}
(\omega^{(\infty)}_{2+}-H_+)|_D&=(\omega^{(\infty)}_2-H)|_D=\frac{-1}{48\pi^2a^2}\,,\notag\\
\partial_t\partial_{t'}(\omega^{(\infty)}_{2+}-H_+)|_D&=\partial_t\partial_{t'}(\omega^{(\infty)}_2-H)|_D
=\frac{1}{480\pi^2a^4}\,.\notag
\end{align}
In the special case $c=0$ the regularised energy density and pressure of the ground state can then be inferred from Equations (\ref{Eqn_00}, \ref{Eqn_trace}), namely
\begin{align}
E^{\mathrm{reg}}(\omega^{(\infty)})&=\frac{1}{480\pi^2a^4}\,,\notag\\
P^{\mathrm{reg}}(\omega^{(\infty)})&=\frac{-m^2}{144\pi^2a^2}-\frac{1}{1440\pi^2a^4}
=\frac{-10m^2a^2-1}{1440\pi^2a^4}\,.\notag
\end{align}

When we drop the simplifying assumption $c=0$, both the Hadamard series and the ground state become more complicated. Note that we can restrict the Hadamard series (\ref{Eqn_Hadseries2}) to $x=x'$, where we have
\begin{align}
H((t,x),(t',x))&=\frac{1}{8\pi^2a^2}\lim_{\epsilon\to0^+}\frac{1}{-\frac12T_{\epsilon}^2}
+\frac{c}{2}\log\left(-\frac12a^2T_{\epsilon}^2\right)
-\frac{c^2}{16}\log\left(-\frac12a^2T_{\epsilon}^2\right)T_{\epsilon}^2+O(T_{\epsilon}^3)\,.\notag
\end{align}
We compare this with the restriction of $\omega_2^{(\infty)}$ in (\ref{eqn:restrictedgroundstate2}), which yields
\begin{align}
(\omega^{(\infty)}_2-H)((t,x),(t',x))&=\frac{-c}{16\pi^2a^2}\log\left(\frac12a^2\right)+\frac{1}{4\pi^2a^2}X_1+O(t-t')
\end{align}
where $X_1$ is given in (\ref{eqn:X1}). Evaluating at $t=t'=0$ gives
\begin{align}
(\omega^{(\infty)}_{2+}-H_+)|_D&=(\omega^{(\infty)}_2-H)|_D\notag\\
&=\frac{-c}{16\pi^2a^2}\log\left(\frac12a^2\right)+\frac{1}{4\pi^2a^2}X_1\,.\label{eqn:restr1}
\end{align}
Similarly, comparing
\begin{align}
\partial_t\partial_{t'}H((t,x),(t',x))&=
\frac{1}{8\pi^2a^4}\lim_{\epsilon\to0^+}\left(\frac{12}{T_{\epsilon}^4}+\frac{c}{T_{\epsilon}^2}
+\frac{c^2}{8}\log\left(-\frac12a^2T_{\epsilon}^2\right)+\frac{3c^2}{8}\right)+O(t-t')\notag
\end{align}
with (\ref{eqn:restrictedgroundstate3}) yields
\begin{align}
\partial_t\partial_{t'}(\omega^{(\infty)}_2-H)((t,x),(t',x))&=
\frac{-c^2}{64\pi^2a^4}\left(3+\log\left(\frac12a^2\right)\right)+\frac{1}{4\pi^2a^4}X_2+O(t-t')\notag
\end{align}
and evaluating at $t=t'$ gives
\begin{align}
\partial_t\partial_{t'}(\omega^{(\infty)}_{2+}-H_+)|_D&=\partial_t\partial_{t'}(\omega^{(\infty)}_2-H)|_D\notag\\
&=\frac{-c^2}{64\pi^2a^4}\left(3+\log\left(\frac12a^2\right)\right)+\frac{1}{4\pi^2a^4}X_2\,.\label{eqn:restr2}
\end{align}
It follows from (\ref{Eqn_00}, \ref{Eqn_trace}) that general ground states have
\begin{align}
E^{\mathrm{reg}}(\omega^{(\infty)})&=
\frac{1}{64\pi^2a^4}\left(16X_2-c^2\log\left(\frac12a^2\right)\right)\,,\notag\\
P^{\mathrm{reg}}(\omega^{(\infty)})&=\frac{1}{192\pi^2a^4}\left(16m^2a^2X_1-16X_2+6c^2
+c(c-4m^2a^2)\log\left(\frac12a^2\right)\right)\,.\label{eqn:EPregground}
\end{align}

\section{The semi-classical Einstein equation}\label{sec_SSE}

In a general spacetime we can use the renormalised stress tensor (\ref{Eqn_Tren}) to write the semi-classical Einstein equation (\ref{Eqn_semicl}) as
\begin{align}\label{Eqn_semicl2}
\frac{1}{\kappa}(G_{ab}+\Lambda g_{ab})&=\langle T^{\mathrm{reg}}_{ab}\rangle_{\omega}+g_{ab}\frac{u_2|_D}{16\pi^2}\notag\\
&+\alpha_1m^4g_{ab}+\alpha_2m^2G_{ab}+\alpha_3I_{ab}+\alpha_4J_{ab}+\alpha_5K_{ab}\\
&+\beta_1\Lambda^2g_{ab}+\beta_2\Lambda G_{ab}+\beta_3\Lambda m^2g_{ab}\,.\notag
\end{align}
Most renormalisation terms on the right-hand side can be absorbed by modifying Newton's constant and the cosmological constant to
\begin{align}
\kappa'&:=\frac{\kappa}{1-\kappa(\alpha_2m^2+\beta_2\Lambda)} \ ,\label{Eqn_kappa'}\\
\Lambda'&:=\frac{\Lambda-\kappa(\alpha_1m^4+\beta_1\Lambda^2+\beta_3\Lambda m^2)}{1-\kappa(\alpha_2m^2+\beta_2\Lambda)} \,. \label{Eqn_Lambda'}
\end{align}
By using $\kappa'$ and $\Lambda'$ instead of $\kappa$ and $\Lambda$ we effectively set $\alpha_1=\alpha_2=\beta_i=0$ for $i=1,2,3$, leading to the following simplified form of the semi-classical Einstein equation:
\begin{align}\label{Eqn_semicl3}
\frac{1}{\kappa'}(G_{ab}+\Lambda' g_{ab})&=\langle T^{\mathrm{reg}}_{ab}\rangle_{\omega}+g_{ab}\frac{u_2|_D}{16\pi^2}\notag\\
&+\alpha_3I_{ab}+\alpha_4J_{ab}+\alpha_5K_{ab} \,.
\end{align}

In an Einstein static universe the right-hand side of (\ref{Eqn_semicl3}) simplifies further, due to Equations (\ref{Eqn_u2simple}) and (\ref{Eqn_IJKsimple}). However, it is no longer justified to absorb the remaining expressions into Newton's constant or the cosmological constant, because the simplifications rely in an essential way on the symmetries of the metric. We therefore write
\begin{align}\label{Eqn_semicl4}
\frac{1}{\kappa'}(G_{ab}+\Lambda' g_{ab})&=\langle T^{\mathrm{reg}}_{ab}\rangle_{\omega}
+g_{ab}\frac{c^2}{32\pi^2a^4}+c'(4RG_{ab}+g_{ab}R^2) \,,
\end{align}
where we introduced
\begin{align}\label{Eqn_c'}
c'(\xi)&=\frac16(3\alpha_3+\alpha_4+\alpha_5) \,.
\end{align}
In a state with a symmetric two-point distribution the expected stress tensor takes the perfect fluid form, so the semi-classical Einstein equation (\ref{Eqn_semicl4}) reduces to the two equations
\begin{align}
E^{\mathrm{reg}}(\omega)+3P^{\mathrm{reg}}(\omega)-\frac{c^2}{8\pi^2a^4}&=
\frac{-1}{\kappa'}(G^a_{\phantom{a}a}+4\Lambda')=\frac{1}{\kappa'}(R-4\Lambda') \,,\label{Eqn_Semicl2}\\
E^{\mathrm{reg}}(\omega)-\frac{c^2}{32\pi^2a^4}+c'R^2&=\frac{1}{\kappa'}(G_{00}-\Lambda')=\frac{1}{2\kappa'}(R-2\Lambda') \,. \label{Eqn_Semicl1}
\end{align}
Note that these are no longer field equations, because both sides of both equations are constant.

We can now formulate our main result:
\begin{theorem}\label{Thm_solns}
Consider an Einstein static universe of radius $a>0$, a cosmological constant $\Lambda\in\mathbb{R}$ and a free scalar quantum field of mass $m\ge0$ and scalar curvature coupling $\xi\in\mathbb{R}$ such that $c:=m^2a^2+6\xi-1>-1$. Fix renormalisation constants $\alpha_1,\ldots\alpha_5,\beta_1,\ldots,\beta_3\in\mathbb{R}$ and define
\begin{align}
Y_1:=&\frac{1}{32\pi^2a^4}\left(-8m^2a^2X_1+c^2+2m^2a^2c\log\left(\frac12a^2\right)\right)
+\frac{1}{\kappa'}(R-4\Lambda')\,, \label{Eqn_defY1}\\
Y_2:=&\frac{1}{64\pi^2a^4}\left(-16X_2+2c^2+c^2\log\left(\frac12a^2\right)\right)
+\frac{1}{2\kappa'}(R-2\Lambda')-c'R^2 \,, \label{Eqn_defY2}
\end{align}
in terms of $R=\frac{6}{a^2}$ and the numbers defined in (\ref{eqn:defX1}), (\ref{eqn:defX2}), (\ref{Eqn_kappa'}), (\ref{Eqn_Lambda'}) and (\ref{Eqn_c'}). Then the set $\mathcal{S}_{qf}$ of symmetric quasi-free states $\omega$ that solve the semi-classical Einstein equation (\ref{Eqn_semicl}) is in bijective correspondence with sequences $\{a_n\}_{n\ge0}\subset\mathbb{R}_{\ge0}$ such that
\begin{align}
m^2\sum_{n=0}^{\infty}a_n&= Y_1 \,,\label{Eqn_Semicl2v2}\\
\sum_{n=0}^{\infty}a_n\frac{l_n^2}{a^2}&=Y_2 \,, \label{Eqn_Semicl1v2}
\end{align}
where the correspondence is given by Theorem \ref{Thm_sym}.
\end{theorem}
\begin{proof*}
We see from (\ref{Eqn_trace}) and (\ref{Eqn_00}) that the two equations (\ref{Eqn_Semicl2}) and (\ref{Eqn_Semicl1}) are equivalent to
\begin{align}
\frac{1}{\kappa'}(R-4\Lambda')&=m^2(\omega_{2+}-H_+)|_D -\frac{c^2}{32\pi^2a^4}\,, \label{Eqn_Semicl2v3}\\
\frac{1}{2\kappa'}(R-2\Lambda')&=\partial_t\partial_{t'}(\omega_{2+}-H_+)|_D+\frac{c^2}{64\pi^2a^4}+c'R^2 \,. \label{Eqn_Semicl1v3}
\end{align}
Using the CCR we have $\omega_{2+}-H_+=(\omega_2-\omega^{(\infty)}_2)+(\omega_{2+}^{(\infty)}-H_+)$, so we obtain from 
(\ref{eqn:restr1}, \ref{eqn:restr2}) that
\begin{align}
(\omega_{2+}-H_+)|_D&=(\omega_2-\omega^{(\infty)}_2)|_D
+\frac{1}{4\pi^2a^2}X_1-\frac{c}{16\pi^2a^2}\log\left(\frac12a^2\right)\notag\\
\partial_t\partial_{t'}(\omega_{2+}-H_+)|_D&=\partial_t\partial_{t'}(\omega_2-\omega^{\infty}_2)|_D
+\frac{1}{4\pi^2a^4}X_2-\frac{c^2}{64\pi^2a^4}\left(3+\log\left(\frac12a^2\right)\right)\,,\notag
\end{align}
which reduces (\ref{Eqn_Semicl2v3}, \ref{Eqn_Semicl1v3}) to
\begin{align}
Y_1&=m^2(\omega_2-\omega^{(\infty)}_2)|_D\notag\\
Y_2&=\partial_t\partial_{t'}(\omega_2-\omega^{\infty}_2)|_D\,.\notag
\end{align}
The right-hand sides of these equations can be rewritten using Theorem \ref{Thm_sym} and the fact that the Gegenbauer polynomials are normalised by $C_n^{(1)}(1)=n+1$, cf.~(\ref{eqn:Gegenbauer1}). The resulting equations are (\ref{Eqn_Semicl1v2}) and (\ref{Eqn_Semicl2v2}). Since quasi-free states are uniquely determined by their two-point distributions this completes the proof.
\end{proof*}

In the massless case, $m=0$, the equations (\ref{Eqn_Semicl2v2}, \ref{Eqn_Semicl1v2}) simplify to
\begin{align}
\frac{R-4\Lambda'}{\kappa'}&=\frac{-c^2}{32\pi^2a^4}\,,\notag\\
\sum_{n=0}^{\infty}a_n\frac{l_n^2}{a^2}&=\frac{1}{64\pi^2a^4}\left(-16X_2+2c^2+c^2\log\left(\frac12a^2\right)
-c^2\frac{R-2\Lambda'}{R-4\Lambda'}\right)-c'R^2\,.\notag
\end{align}
Note that the first of these equations is now independent of the choice of state, so it imposes a relation between the parameters of the theory with the renormalisation constants. E.g., if $\kappa'>0$ and $\Lambda'<\frac14R$ there are no solutions.

Similarly, in the special case $c=0$ we have $l_n=n+1$ and the expressions for $X_1$, $X_2$, $Y_1$ and $Y_2$ simplify, cf.~(\ref{eqn:X1}, \ref{eqn:X2}). In this case the semi-classical Einstein equation becomes 
\begin{align}
m^2\sum_{n=0}^{\infty}a_n&=Y_1
=\frac{m^2}{48\pi^2a^2}+\frac{1}{\kappa'}(R-4\Lambda'),\notag\\
\sum_{n=0}^{\infty}a_n\frac{(n+1)^2}{a^2}&=Y_2
=\frac{-1}{480\pi^2a^2}+\frac{1}{2\kappa'}(R-2\Lambda')-c'R^2\,.\notag
\end{align}
The equations become particularly simple when both $m=0$ and $c=0$, i.e.~in the massless conformally coupled case.

In the general case the numbers $Y_1$ and $Y_2$ in Theorem \ref{Thm_solns} are less explicit, because the expressions for $X_1$ and $X_2$ involve an infinite series. Nevertheless, the complete set of symmetric quasi-free solutions to the semi-classical Einstein equation is in principle fully parameterised by Theorem \ref{Thm_solns} in combination with Theorem \ref{Thm_sym}.

The following result about solutions which are not necessarily quasi-free is a consequence of Theorem \ref{Thm_solns}.
\begin{corollary}\label{cor:solns}
Under the assumptions of Theorem \ref{Thm_solns} let $\mathcal{S}$ denote the set of all (not necessarily quasi-free) solutions with a symmetric two-point distribution. Then the following statements are true:
\begin{enumerate}
\item $\mathcal{S}$ contains the ground state iff $Y_1=Y_2=0$.
\item If $m=0$, then $\mathcal{S}\not=\emptyset$ if and only if $Y_2\ge Y_1=0$.
\item If $m>0$, then $\mathcal{S}\not=\emptyset$ if and only if $Y_1=Y_2=0$ or $m^2Y_2\ge (m^2+\xi R)Y_1>0$.
\item If $\mathcal{S}$ contains more than one solution, then it contains infinitely many solutions.
\item $\mathcal{S}$ consists of a unique solution iff $\mathcal{S}$ contains the ground state.
\end{enumerate}
\end{corollary}
\begin{proof*}
We first prove the first four statements for the set $\mathcal{S}_{qf}$ of quasi-free states, instead of $\mathcal{S}$. The first two statements are easy consequences of the fact that $a_n\ge0$ for all $n\ge0$ with $a_n=0$ for all $n$ only in the ground state.

For the third statement we use the fact that $l_n^2\ge l_0^2=c+1=a^2(m^2+\xi R)$ with strict inequality when $n>0$, so any solution must have $m^2Y_2\ge (m^2+\xi R)Y_1$. If $Y_1=0$, then $a_n=0$ for all $n\ge0$ (since $m>0$) and hence $Y_2=0$ too. If $Y_1>0$, then $(m^2+\xi R)Y_1>0$, since we assumed $m^2+\xi R>0$ to guarantee the existence of a ground state. Conversely, if $Y_1=Y_2=0$ then $\mathcal{S}_{qf}$ contains the ground state. If $m^2Y_2\ge (m^2+\xi R)Y_1>0$ we construct a solution as follows. We can pick $N\in\mathbb{N}$ sufficiently large to ensure that $\frac{l_N^2Y_1}{a^2m^2}\ge Y_2$. We set $a_n:=0$ for $n\not\in\{0,N\}$ and
\begin{align}
a_0&:=\frac{a^2}{l_N^2-l_0^2}\left(\frac{l_N^2Y_1}{a^2m^2}-Y_2\right)\,,\notag\\
a_N&:=\frac{a^2}{l_N^2-l_0^2}\left(Y_2-\frac{l_0^2}{a^2m^2}Y_1\right)\,.\notag
\end{align}
We note that $a_0\ge0$, $a_N\ge0$ and these values of $a_n$ yield a solution. This proves the third statement with $\mathcal{S}_{qf}$ instead of $\mathcal{S}$.

The construction of a solution above works for all sufficiently large $N$ and it yields distinct solutions when $m^2Y_2>(m^2+\xi R)Y_1>0$. Thus we see that $\mathcal{S}_{qf}$ contains infinitely many solutions when $m>0$ and $m^2Y_2> (m^2+\xi R)Y_1>0$. On the other hand, when $m>0$ and $m^2Y_2=(m^2+\xi R)Y_1$, then every solution must have $a_n=0$ for all $n>0$ and there can be only one solution. Similarly, if $m=0$ and $Y_2>Y_1=0$, then $\mathcal{S}_{qf}$ contains infinitely many solutions, because we can choose $a_n=0$ for all but one $n$. On the other hand, if $Y_2=Y_1=0$, then $\mathcal{S}_{qf}$ contains only one solution, namely the ground state. This proves the fourth statement with $\mathcal{S}_{qf}$ instead of $\mathcal{S}$
and the following modification of the last statement:
\begin{enumerate}
\item[5'.] $\mathcal{S}_{qf}$ consists of a unique solution iff $\mathcal{S}_{qf}$ contains the ground state or
$m^2Y_2=(m^2+\xi R)Y_1>0$.
\end{enumerate}

To deduce the desired statements for $\mathcal{S}$ we note that the ground state is quasi-free and $\mathcal{S}\supset \mathcal{S}_{qf}$, so
$\mathcal{S}$ contains the ground state iff $\mathcal{S}_{qf}$ contains it. Furthermore, if $\omega\in\mathcal{S}$, then $\omega$ must have a two-point distribution $\omega_2$. Because the equations in Theorem \ref{Thm_solns} only depend on $\omega_2$, the quasi-free state with the same two-point distribution $\omega_2$ must be in $\mathcal{S}_{qf}$. Hence $\mathcal{S}_{qf}=\emptyset$ iff $\mathcal{S}=\emptyset$. This proves the first three items for $\mathcal{S}$.

If $\mathcal{S}$ contains the ground state, then $\mathcal{S}_{qf}$ contains only the ground state by 5'. All solutions in $\mathcal{S}$ must then have the same two-point distribution as the ground state. Because the ground state is pure, we can apply a theorem of Kay \cite{Kay1993} to see that there is no other state with the same two-point distribution. Hence, $\mathcal{S}$ contains a unique element.

Conversely, if $\mathcal{S}$ contains a unique solution $\omega$, then $\mathcal{S}_{qf}$ cannot be empty and we must have $\mathcal{S}_{qf}=\mathcal{S}$ containing the same unique solution. If this is the ground state, then the fifth statement holds. Otherwise we must have $m>0$ and $m^2Y_2=(m^2+\xi R)Y_1>0$ by 5' and therefore $a_n=0$ for all $n>0$. Using Theorem \ref{Thm_sym} we then find that
\begin{align}
\omega_2(x,x')&=\omega_2^{(\infty)}(x,x')+C\cos(\gamma(t-t'))\notag
\end{align}
where $\gamma:=\frac{\sqrt{c+1}}{a}=\sqrt{m^2+\xi R}$ and $C>0$ is a constant. We now obtain a contradiction, because there are infinitely many states with this two-point distribution, given by\footnote{Note that the one-point distribution $\omega_1$ of these solutions does not respect all symmetries of the spacetime, because it is not constant in time. In the smaller set of states for which also $\omega_1$ respects the spacetime symmetries, there might be a unique solution which is not the ground state. However, it is unclear why one would impose such an additional restriction on $\omega_1$, because unlike the symmetry of $\omega_2$, the symmetry of $\omega_1$ is unnecessary to simplify the analysis.}
\begin{align}
\omega(W(f))&=e^{i\omega_1(f)-\frac12w_2(f,f)}\omega^{(\infty)}(W(f))\,,\notag
\end{align}
where the one-point function is $\omega_1(t,x)=\lambda\sin(\gamma t)$ with $\lambda\in[-\sqrt{C},\sqrt{C}]$ and
\begin{align}
w_2((t,x),(t',x'))&=C\cos(\gamma(t-t'))-\omega_1(t,x)\omega_1(t',x')\notag\\
&=C\cos(\gamma t)\cos(\gamma t')+(C-\lambda^2)\sin(\gamma t)\sin(\gamma t')\,.\notag
\end{align}
This definition of $\omega$ is positive, because $C>0$ and $C-\lambda^2\ge0$, so $w_2$ is of positive type. Note that we may even obtain infinitely many solutions with a vanishing one-point function by considering a mixture of the states above with $+\lambda$ and $-\lambda$, leading to
\begin{align}
\omega(W(f))&=\cos\left(\omega_1(f)\right)e^{-\frac12w_2(f,f)}\omega^{(\infty)}(W(f))\,.\notag
\end{align}
(These states differ in $\omega_4$ as $\lambda$ varies.) This contradiction shows that when $\mathcal{S}$ contains a unique state, it must be the ground state.

Finally, let us suppose that $\mathcal{S}$ contains more than one state. If $\mathcal{S}_{qf}$ also contains more than one state, then it contains infinitely many and hence so does $\mathcal{S}$. Alternatively, $\mathcal{S}_{qf}$ could contain only one state.
This cannot be the ground state, because then $\mathcal{S}$ would only contain one state. We must therefore have $m^2Y_2=(m^2+\xi R)Y_1>0$ and we have just seen that $\mathcal{S}$ contains infinitely many states in this case.
\end{proof*}

\begin{remark}\label{rem:scale}
Recall that the renormalisation constants $\alpha_1,\ldots,\alpha_5$ and $\beta_1,\ldots,\beta_3$ in (\ref{Eqn_Tren}) are functions of $\xi$ only. If we vary $a>0$ while keeping the dimensionless quantities $\xi$, $m^2a^2$, $\Lambda a^2$ and $\frac{\kappa}{a^2}$ fixed, then $c$ remains constant by its definition (\ref{eqn:defc}) and we see from (\ref{Eqn_defY1}, \ref{Eqn_defY2}, \ref{eqn:X1}, \ref{eqn:X2}) that all terms in $Y_1$ and $Y_2$ scale like $a^{-4}$, except for the terms containing $\log\left(\frac12a^2\right)$. This means that the sets of solutions at different values of $a$ are related in a non-trivial way. E.g., if for some choice of the parameters we have $Y_1$ and $Y_2$ which do not admit solutions, then we can change $a$ to $\lambda a$ with $\lambda>0$ to find new values $\tilde{Y}_1$ and $\tilde{Y}_2$ with
\begin{align}
\tilde{Y}_1&=\frac{1}{\lambda^4}\left(Y_1+ \frac{m^2a^2c}{8\pi^2a^4}\log(\lambda)\right)\,,\notag\\
\tilde{Y}_2&=\frac{1}{\lambda^4}\left(Y_2+ \frac{c^2}{32\pi^2a^4}\log(\lambda)\right)\,.\notag
\end{align}
In order to find solutions we would like to have $m^2a^2\tilde{Y}_2\ge(c+1)\tilde{Y}_1>0$, which gives after a little algebra
\begin{align}
-\frac14Y_1&<\frac{m^2a^2c}{32\pi^2a^4}\log(\lambda)\le\frac{m^2a^2Y_2-(c+1)Y_1}{3c+4}\,.\notag
\end{align}
We can find $\lambda$ satisfying these inequalities when $m\not=0$, $c\not=0$ and $m^2a^2Y_2>\frac{c}{4}Y_1$. However, we do not need $Y_1>0$ or $m^2a^2Y_2\ge (c+1)Y_1$.
\end{remark}

One of the main difficulties in studying the semi-classical Einstein equation is the ambiguity that arises from the renormalisation freedom. This is illustrated in the following result.
\begin{proposition}
For any allowed choice of $(a,\Lambda,m,\xi)$ in Theorem \ref{Thm_solns} with $m\not=0$ or $\Lambda\not=0$ we can obtain any of the three alternatives $\mathcal{S}=\emptyset$, or $\mathcal{S}=\{\omega^{(\infty)}\}$, or $\mathcal{S}$ contains infinitely many solutions, by choosing appropriate renormalisation constants $\alpha_1,\ldots\alpha_5,\beta_1,\ldots,\beta_3\in\mathbb{R}$.
\end{proposition}
\begin{proof*}
We see from (\ref{Eqn_kappa'}, \ref{Eqn_Lambda'}, \ref{Eqn_c'}) that $\kappa'$ only depends on the renormalisation constants
$\alpha_2$ and $\beta_2$, $\frac{\kappa'}{\Lambda'}$ only depends on $\alpha_1$, $\beta_1$ and $\beta_3$ and $c'$ only depends on $\alpha_3$, $\alpha_4$ and $\alpha_5$. This means that we can choose $\kappa'$, $\Lambda'$ and $c'$ independently and as long as $m\not=0$ or $\Lambda\not=0$ we also see from the same equations that we can set them equal to arbitrary real numbers. This allows us to give $Y_1$ and $Y_2$ in Theorem \ref{Thm_solns} arbitrary values and the result then follows from Corollary \ref{cor:solns}.
\end{proof*}
When $m=\Lambda=0$ the proof of this proposition fails, because the renormalisation freedom can no longer affect $\kappa$ or $\Lambda=0$. We can still choose $c'$ freely and ensure that $Y_2<0$ and $\mathcal{S}=\emptyset$, but the value of $Y_1$ is entirely determined by the parameters $a$ and $\xi$ and it can prevent us from achieving $\mathcal{S}\not=\emptyset$.

\section{Properties of solutions}\label{sec_properties}

In general one expects some of the solutions to the semi-classical Einstein equation to be unphysical, e.g.~so-called runaway solutions. Following \cite{Parker+1993} one may use the reduction of order formalism to restrict attention to those solutions whose dependence on $\hbar$ is of a perturbative nature, which is in line with the range of the validity of the equation itself. We will now investigate which of the solutions we found satisfy this additional condition. In order to do this, we will first restore powers of $\hbar$ and review the reduction of order formalism in general spacetimes.

Whereas the metric is classical, the anti-symmetric part of the two-point function $\omega_{2-}$ clearly scales linearly with $\hbar$. More generally, we will decompose the two-point distribution as
\begin{align}
\omega_2(x,x')&=\omega_1(x)\omega_1(x')+\hbar\,\omega_2^T(x,x')\,,\label{Eqn_omega2T}
\end{align}
where we treat the one-point function $\omega_1$ as a classical solution to the Klein-Gordon equation and the truncated two-point distribution $\omega_2^T$ 
receives a factor $\hbar$. Similarly, the terms in the Hadamard parametrix and in the renormalisation freedom of the stress tensor also receive a factor $\hbar$.
Using Equations (\ref{Eqn_omega2T},\ref{Eqn_Tren},\ref{Eqn_Treg}) we can then decompose the renormalised stress tensor as
\begin{align}
\langle T^{\mathrm{ren}}_{ab}\rangle_{\omega}&=T_{ab}\left[\omega_1\right]+\hbar\,\langle T^{\mathrm{ren}}_{ab}\rangle_{\omega^T}
\,,\notag
\end{align}
where the first term is the (classical) stress tensor of the classical Klein-Gordon solution $\omega_1$ and the second term is computed using $\omega_2^T$ instead of $\omega_2$. (There is a slight abuse of notation here, because $\omega^T$ does not denote a quantum state, but the expression only depends on $\omega_2^T$ and is therefore unambiguous.) We then consider the semi-classical Einstein equation to be valid up to order $\hbar^2$, i.e. (\ref{Eqn_semicl3}) becomes
\begin{align}\label{Eqn_semicl5}
\frac{1}{\kappa'}(G_{ab}+\Lambda' g_{ab})&=T_{ab}[\omega_1]+\hbar\,\langle T^{\mathrm{reg}}_{ab}\rangle_{\omega^T}
+\hbar\,g_{ab}\frac{u_2|_D}{16\pi^2}+\hbar\,\alpha_3I_{ab}+\hbar\,\alpha_4J_{ab}+\hbar\,\alpha_5K_{ab} +O(\hbar^2)\,,
\end{align}
where the expressions (\ref{Eqn_kappa'},\ref{Eqn_Lambda'}) for $\kappa'$ and $\Lambda'$ should be modified by introducing a factor $\hbar$ for each of the coefficients $\alpha_i,\beta_i$, so $\kappa'=\kappa+O(\hbar)$ and $\Lambda'=\Lambda+O(\hbar)$. (We will keep the primes on $\kappa'$ and $\Lambda'$ to remind the reader that some of the renormalisation freedom has been absorbed here.)

The reduction of order scheme now modifies Equation (\ref{Eqn_semicl5}) by terms of order $\hbar^2$ in order to remove terms that depend on derivatives of the metric of order $\ge3$. Indeed, any solution to (\ref{Eqn_semicl5}) also satisfies
\begin{align}
\frac{1}{\kappa'}(G_{ab}+\Lambda' g_{ab})&=T_{ab}[\omega_1]+O(\hbar)\,.\label{Eqn_semicl6}
\end{align}
This can be used to remove the higher derivative terms in $I_{ab}$, $J_{ab}$ and $K_{ab}$. Indeed, taking derivatives of (\ref{Eqn_semicl6}) and a trace we find
\begin{align}
\nabla_c\nabla_dR_{ab}-\frac12g_{ab}\nabla_c\nabla_dR&=\kappa'\nabla_c\nabla_dT_{ab}[\omega_1]+O(\hbar)\,,\notag\\
-\nabla_c\nabla_dR&=\kappa'\nabla_c\nabla_dT^a_{\phantom{a}a}[\omega_1]+O(\hbar)\,,\notag
\end{align}
and substituting the second line in the first gives
\begin{align}
\nabla_c\nabla_dR_{ab}&=\kappa'\nabla_c\nabla_dT_{ab}[\omega_1]-\frac12\kappa'g_{ab}\nabla_c\nabla_dT^e_{\phantom{e}e}[\omega_1]
+O(\hbar)\,.\label{Eqn_reducedR}
\end{align}
Inserting various contractions of this into (\ref{Eqn_I},\ref{Eqn_J},\ref{Eqn_K}) yields
\begin{align}
I_{ab}&=2RR_{ab}-\frac12g_{ab}R^2
+2\kappa'\nabla_a\nabla_bT^c_{\phantom{c}c}[\omega_1]
-2\kappa'g_{ab}\Box T^c_{\phantom{c}c}[\omega_1]+O(\hbar)\,,\notag\\
J_{ab}&=2R_{acbd}R^{cd}-\frac12g_{ab}R_{cd}R^{cd}-\kappa'g_{ab}\Box T^c_{\phantom{c}c}[\omega_1]+\kappa'\Box T_{ab}[\omega_1]
+\kappa'\nabla_a\nabla_bT^c_{\phantom{c}c}[\omega_1]+O(\hbar)\,,\notag\\
K_{ab}&=2R_{acde}R_b^{\phantom{b}cde}+4R_{acbd}R^{cd}-4R_{ac}R^c_{\phantom{c}b}-\frac12g_{ab}R_{cdef}R^{cdef}\notag\\
&\quad +4\kappa'\Box T_{ab}[\omega_1]-2\kappa'g_{ab}\Box T^e_{\phantom{e}e}[\omega_1]
+2\kappa'\nabla_a\nabla_bT^c_{\phantom{c}c}[\omega_1]+O(\hbar)\,.\notag
\end{align}
These expressions can now be substituted into (\ref{Eqn_semicl5}) to obtain an equation that no longer depends on higher derivatives of the metric. The Klein-Gordon equation for the $n$-point distributions remains unmodified.

We will not investigate the order-reduced semi-classical Einstein equation in general, but only make some comments for the Einstein symmetric universe, where things simplify considerably. For a state with $\omega_1=0$ Equation (\ref{Eqn_reducedR}) shows that the reduction of order simply means that all terms in $I_{ab}$, $J_{ab}$ and $K_{ab}$ involving derivatives of $R_{ab}$ are dropped. However, in the Einstein static universe one can also use the spacetime symmetries to see that $\nabla_cR_{ab}=0$, so when $\omega_1=0$ the reduced order equation coincides with the original semi-classical Einstein equation. In particular, all solutions in $\mathcal{S}_{qf}$ are also solutions to the reduced order equation. This is perhaps not surprising: static solutions do not exhibit the kind of runaway behaviour that the reduction of order formalism is supposed to eradicate.

In contrast, the set $\mathcal{S}$ may contain solutions of the semi-classical Einstein equation with non-vanishing $\omega_1$, for which the reduced form of the semi-classical Einstein equation differs from the original equation and may be violated. E.g., we may have the spatially homogeneous Klein-Gordon solution $\omega_1(t,x)=\sin\left(t\sqrt{m^2+\xi R}\right)$ (cf.~the proof of Corollary \ref{cor:solns}). Although the oscillating behaviour of $\omega_1$ does not exhibit runaway behaviour, it does violate Equation (\ref{Eqn_semicl6}) in the limit $\hbar\to0$. We will not investigate whether $\omega_1$ could be combined with a different metric to yield a solution to the reduced order equation. We do wish to emphasise, however, that the result of Corollary \ref{cor:solns} in Section \ref{sec_SSE} remains valid also for the smaller set of solutions of the reduced order equation, because the proof only relies on states in $\mathcal{S}$ with a vanishing one-point distribution.

In the remainder of this section we will investigate some further properties of the solutions in $\mathcal{S}_{qf}$. Before we proceed, however, it will be useful to consider the ground state representation, which is a bosonic Fock space $\mathcal{H}=\mathcal{F}_+(\mathcal{H}_1)$ over the one-particle Hilbert space $\mathcal{H}_1$. We can identify $\mathcal{H}_1=L^2(\mathbb{S}^3)$ and for any $f_1,f_2\in C_0^{\infty}(M)$ we have
\begin{align}
\omega_2^{(\infty)}(\overline{f_1}a^3\sqrt{\det h},f_2a^3\sqrt{\det h})&=
\frac12\langle K(f_1),K(f_2)\rangle_{L^2(\mathbb{S}^3)}\notag
\end{align}
in terms of  the one-particle structure $K:C_0^{\infty}(M)\to L^2(\mathbb{S}^3)$, which is given by
\begin{align}
K(f)&:=a^2\left(A^{-\frac14}\partial_tGf|_{t=0}-\frac{i}{a}A^{+\frac14}Gf|_{t=0}\right)\,,\label{eqn:Kstructure}
\end{align}
where $G:=G^--G^+$ is the difference of the advanced and retarded fundamental solutions for the Klein-Gordon equation
(\ref{Eqn_clKG}). (This formula can be deduced from (\ref{Eqn_ground2}).)

We can describe many other quasi-free states using density matrices in $\mathcal{H}$ as follows. Suppose that $H_1\ge0$ is a self-adjoint operator on $L^2(\mathbb{S}^3)$. We decompose $L^2(\mathbb{S}^3)=\mathrm{ker}(H_1)\oplus \mathrm{ker}(H_1)^{\perp}$, so that $\mathcal{H}=\mathcal{F}_+(\mathrm{ker}(H_1))\otimes \mathcal{F}_+(\mathrm{ker}(H_1)^{\perp})$, cf.~Sec.~3.3.7 of \cite{Derezinski+2013}. We let $P_0$ denote the orthogonal projection in $\mathcal{F}_+(\mathrm{ker}(H_1))$ onto the Fock vacuum. Suppose that $e^{-H_1}>0$ is a trace class operator on $\mathrm{ker}(H_1)^{\perp}$ and let $\{v_j\}_{j=1}^{\infty}$ be an orthonormal eigenbasis for $e^{-H_1}$ with corresponding eigenvalues $e^{-\epsilon_j}>0$, so that $H_1$ has eigenvalues $\epsilon_j>0$. The second quantisation of $H_1$ from $\mathrm{ker}(H_1)^{\perp}$ to $\mathcal{F}_+(\mathrm{ker}(H_1)^{\perp})$ is the (unbounded) operator $H$ given by
\begin{align}
H&:=\sum_{j=1}^{\infty}\epsilon_ja^*(v_j)a(v_j)\notag
\end{align}
in terms of creation and annihilation operators. $e^{-H}$ is a trace class operator on $\mathcal{F}_+(\mathrm{ker}(H_1)^{\perp})$ and hence $P_0\otimes e^{-H}$ is a trace class operator on $\mathcal{H}$. The density matrix
\begin{align}
\rho_{H_1}&:=\frac{1}{\mathrm{Tr}_{\mathcal{H}}(P_0\otimes e^{-H})}P_0\otimes e^{-H}\notag
\end{align}
defines a quasi-free state by $\omega^{(H_1)}(A):=\mathrm{Tr}_{\mathcal{H}}(\rho_{H_1}A)$ with 
\begin{align}
\omega^{(H_1)}_2(fa^3\sqrt{\det h},fa^3\sqrt{\det h})&=
\frac12\left\langle K(f), \left(Q_0 + Q^{\perp}\frac{I+e^{-H_1}}{I-e^{-H_1}}\right)K(f)\right\rangle_{L^2(\mathbb{S}^3)}
\label{eqn:normalqf}
\end{align}
for all $f\in C_0^{\infty}(M,\mathbb{R})$, where $Q_0$ is the orthogonal projection in $L^2(\mathbb{S}^3)$ onto $\mathrm{ker}(H_1)$ and $Q^{\perp}=I-Q_0$. We refer to \cite{Bratteli+} Proposition 5.2.27 and 5.2.28 for detailed computations in the case where $Q_0=0$ and we note that the general case easily follows from the tensor product structure of the Fock space.
This construction applies in particular to $\beta$-KMS states with $\beta>0$ and $H_1=\frac{\beta}{a}\sqrt{A}$, where $\frac{1}{a}\sqrt{A}$ is the one-particle Hamiltonian, cf.~(\ref{Eqn_KMS2}).

\begin{proposition}\label{prop:normal}
Under the assumptions of Theorem \ref{Thm_solns}, every state $\omega\in\mathcal{S}_{qf}$ is of the form $\omega(A)=\mathrm{Tr}_{\mathcal{H}}(\rho_{H_1}A)$, where (\ref{eqn:normalqf}) holds with
\begin{align}
H_1&=\sum_{n=0,\ a_n\not=0}^{\infty}\log\left(\frac{(n+1)^2+2\pi^2a^2l_na_n}{2\pi^2a^2l_na_n}\right)E_n^{(3)}\,,\notag
\end{align}
where the sum runs over all $n\in\mathbb{N}_0$ with $a_n\not=0$.
\end{proposition}
\begin{proof*}
Because $\mathcal{S}_{qf}$ only contains quasi-free and symmetric solutions, we know that the two-point distribution is of the form given in Theorem \ref{Thm_sym}. For any $f\in C_0^{\infty}(M,\mathbb{R})$ this can be written as
\begin{align}
\omega_2(fa^3\sqrt{\det h},fa^3\sqrt{\det h})&=\frac12\left\langle K(f),\left(I+
\sum_{n=0}^{\infty}\frac{4\pi^2a^2l_na_n}{(n+1)^2}E_n^{(3)}\right)K(f)\right\rangle_{L^2(\mathbb{S}^3)}\,,\notag
\end{align}
where we used Equation (\ref{eqn:projectionkernel}). This is of the form (\ref{eqn:normalqf}), when we have $e^{-H_1}=\sum_{n=0}^{\infty}x_nE_n^{(3)}$ with $x_n=1$ when $a_n=0$ and
\begin{align}
\frac{1+x_n}{1-x_n}&=1+\frac{4\pi^2a^2l_na_n}{(n+1)^2}\notag
\end{align}
otherwise. Solving for $H_1$ leads to the claimed result. To find a density matrix we need $e^{-H_1}Q^{\perp}$ to be a trace class operator. Note that the projections  $E_n^{(3)}$ project onto the linear subspaces $L_n^{(3)}$ of dimension $(n+1)^2$, so
\begin{align}
\mathrm{Tr}_{L^2(\mathbb{S}^3)}(e^{-H_1}Q^{\perp})&=
\sum_{n=0}^{\infty}\frac{2\pi^2a^2l_na_n}{(n+1)^2+2\pi^2a^2l_na_n}(n+1)^2\notag\\
&\le 2\pi^2a^2\sum_{n=0}^{\infty}l_na_n\,,\notag
\end{align}
which is finite when $\omega\in\mathcal{S}_{qf}$, due to (\ref{Eqn_Semicl1v2}).
\end{proof*}

When $\mathcal{S}_{qf}\not=\emptyset$ we can find solutions with nice additional properties: 
\begin{proposition}\label{prop:specialsolns}
Under the assumptions of Theorem \ref{Thm_solns} the following are true:
\begin{enumerate}
\item $\mathcal{S}_{qf}$ contains at most one KMS-state.
\item If $\mathcal{S}_{qf}\not=\emptyset$, then it contains a unique state $\omega$ which minimises the von Neumann entropy
$S_{\mathrm{vN}}(\omega)=\mathrm{Tr}_{\mathcal{H}}(\rho_{H_1}\log(\rho_{H_1}))$, where $H_1$ is as in Proposition \ref{prop:normal}. When $m=0$ this state is a $\beta$-KMS state for some $\beta>0$.
\end{enumerate}
\end{proposition}
\begin{proof*}
From (\ref{eqn_thermalAn}) we see that KMS states have
\begin{align}
\sum_{n=0}^{\infty}a_n^{(\beta)}\frac{l_n^2}{a^2}&=
\sum_{n=0}^{\infty}\frac{(n+1)^2}{2\pi^2 a^4}\ l_n\frac{e^{-\beta\frac{l_n}{a}}}{1-e^{-\beta\frac{l_n}{a}}}\,,\notag
\end{align}
which is a strictly decreasing function in $\beta>0$, so there can be at most one value of $\beta$ where the sum equals $Y_2$. In view of Theorem \ref{Thm_solns} this means that there is at most one KMS state in $\mathcal{S}_{qf}$.

Using $\mathcal{H}=\mathcal{F}_+(\mathrm{ker}(H_1))\otimes \mathcal{F}_+(\mathrm{ker}(H_1)^{\perp})$ we can write the von Neumann entropy of $\omega\in\mathcal{S}_{qf}$ as
\begin{align}
\mathrm{Tr}_{\mathcal{H}}(\rho_{H_1}\log(\rho_{H_1}))&=\frac{1}{\mathrm{Tr}_{\mathcal{H}}(P_0\otimes e^{-H})}
\mathrm{Tr}_{\mathcal{H}}(P_0\otimes e^{-H}(\log(P_0)\otimes I -I\otimes H))\notag\\
&\quad -\log(\mathrm{Tr}_{\mathcal{H}}(P_0\otimes e^{-H}))\notag\\
&=\frac{-1}{\mathrm{Tr}_{\mathcal{H}}(P_0\otimes e^{-H})}\mathrm{Tr}_{\mathcal{H}}(P_0\otimes He^{-H})
-\log(\mathrm{Tr}_{\mathcal{H}}(P_0\otimes e^{-H}))\,.\notag
\end{align}
An eigenvector $v_j$ of $h$ with eigenvalue $\epsilon_j>0$ generates a Fock space $\mathcal{F}_+(\mathbb{C}v_j)$ where
\begin{align}
\mathrm{Tr}_{\mathcal{F}_+(\mathbb{C}v_j)}(e^{-H})&=\sum_{n=0}^{\infty}e^{-n\epsilon_j}=\frac{1}{1-e^{-\epsilon_j}}
\,,\notag\\
\mathrm{Tr}_{\mathcal{F}_+(\mathbb{C}v_j)}(He^{-H})&=\sum_{n=0}^{\infty}n\epsilon_je^{-n\epsilon_j}
=-\epsilon_j\partial_{\epsilon_j}\sum_{n=0}^{\infty}e^{-n\epsilon_j}
=-\epsilon_j\partial_{\epsilon_j}\frac{1}{1-e^{-\epsilon_j}}
=\frac{\epsilon_je^{-\epsilon_j}}{(1-e^{-\epsilon_j})^2}\,.\notag
\end{align}
Because $\mathcal{F}_+(\mathrm{ker}(H_1)^{\perp})$ is a tensor product of such Fock spaces for all $v_j$ we have
\begin{align}
\mathrm{Tr}_{\mathcal{H}}(P_0\otimes e^{-H})&=\prod_{j=1}^{\infty}\frac{1}{1-e^{-\epsilon_j}}\,,\notag\\
\mathrm{Tr}_{\mathcal{H}}(P_0\otimes \epsilon_ja^*(v_j)a(v_j)e^{-H})&=
\frac{\epsilon_je^{-\epsilon_j}}{1-e^{-\epsilon_j}}\mathrm{Tr}_{\mathcal{H}}(P_0\otimes e^{-H})\notag
\end{align}
and therefore
\begin{align}
\mathrm{Tr}_{\mathcal{H}}(\rho_{H_1}\log(\rho_{H_1}))&=-\sum_{j=1}^{\infty}\frac{\epsilon_je^{-\epsilon_j}}{1-e^{-\epsilon_j}}
+\sum_{j=1}^{\infty}\log(1-e^{-\epsilon_j})\,.\notag
\end{align}
Reading off the values $\epsilon_j$ from Proposition \ref{prop:normal} we find in terms of the $a_n$
\begin{align}
\mathrm{Tr}_{\mathcal{H}}(\rho_{H_1}\log(\rho_{H_1}))&=
\sum_{n=0,\ a_n\not=0}^{\infty}(n+1)^2\log\left(\frac{(n+1)^2}{(n+1)^2+2\pi^2a^2l_na_n}\right)\notag\\
&\quad+2\pi^2a^2l_na_n\log\left(\frac{2\pi^2a^2l_na_n}{(n+1)^2+2\pi^2a^2l_na_n}\right)\notag\\
&=\sum_{n=0}^{\infty}2\pi^2a^2l_na_n\log\left(\frac{2\pi^2a^2l_na_n}{(n+1)^2}\right)
-((n+1)^2+2\pi^2a^2l_na_n)\log\left(1+\frac{2\pi^2a^2l_na_n}{(n+1)^2}\right)
\,,\notag
\end{align}
where the overall factor $(n+1)^2$ accounts for the multiplicities of the eigenvalues of $A$ and the terms with $a_n=0$ do not contribute to the sum.

The second statement is trivial when $\mathcal{S}_{qf}$ contains one element. When $\mathcal{S}_{qf}$ contains more than one element we can minimise the function
\begin{align}
F(\{a_n\},\lambda_1,\lambda_2)&:=S_{\mathrm{vN}}(\rho_{H_1})-\lambda\left(Y_1-m^2\sum_{n=0}^{\infty}a_n\right)
-2\pi^2a^3\beta\left(Y_2-\sum_{n=0}^{\infty}\frac{l_n^2}{a^2}a_n\right)\,,\notag
\end{align}
where the Lagrange multipliers $\lambda$ and $\beta$ are used to enforce (\ref{Eqn_Semicl2v2},\ref{Eqn_Semicl1v2}). We then find that any minimum must be of the form
\begin{align}
a_n&=\frac{(n+1)^2}{2\pi^2a^2l_n}\frac{1}{\exp\left(\frac{\lambda m^2}{2\pi^2a^2l_n}+\frac{\beta}{a}l_n\right)-1}
\label{eqn:an}
\end{align}
where $\lambda,\beta>0$ need to be chosen to guarantee (\ref{Eqn_Semicl2v2},\ref{Eqn_Semicl1v2}). In order to have $a_n\ge0$ 
and to have converging sums in (\ref{Eqn_Semicl2v2},\ref{Eqn_Semicl1v2}) we must have $\beta>0$ and
$\lambda m^2>-2\pi^2al_0^2\beta$. Because the coefficients are strictly monotonically increasing functions of $\beta>0$ and of $\lambda>-\frac{2\pi^2al_0^2}{m^2}\beta$ when $m>0$, there exists at most one set of solutions $a_n$ of (\ref{Eqn_Semicl2v2},\ref{Eqn_Semicl1v2}) of the form (\ref{eqn:an}).

To show that a choice for $\beta$ and $\lambda$ exists we first consider the case $m>0$. We may assume that $m^2Y_2>(m^2+\xi R)Y_1$, otherwise there would only be one solution. We then note that the sums
\begin{align}
G_1(\lambda,\beta)&:=m^2\sum_{n=0}^{\infty}a_n
=\sum_{n=0}^{\infty}\frac{(n+1)^2m^2}{2\pi^2a^2l_n}\frac{1}{\exp\left(\frac{\lambda m^2}{2\pi^2a^2l_n}+\frac{\beta}{a}l_n\right)-1}\,,\notag\\
G_2(\lambda,\beta)&:=\sum_{n=0}^{\infty}\frac{l_n^2}{a^2}a_n
=\sum_{n=0}^{\infty}\frac{(n+1)^2l_n}{2\pi^2a^4}\frac{1}{\exp\left(\frac{\lambda m^2}{2\pi^2a^2l_n}+\frac{\beta}{a}l_n\right)-1}\notag
\end{align}
are differentiable functions of $\beta>0$ and $\lambda>\lambda_0(\beta):=-\frac{2\pi^2al_0^2}{m^2}\beta$ and they are both strictly monotonically decreasing in $\beta$ and $\lambda$. Note that for all $\beta>0$ and $\lambda>\lambda_0(\beta)$ and for all $n\ge0$
\begin{align}
0&<a_n\le \frac{(n+1)^2}{2\pi^2a^2l_n}\frac{\exp\left(-\frac{\lambda m^2}{4\pi^2a^2l_n}-\frac{\beta}{2a}l_n\right)}
{2\sinh\left(\sqrt{\frac{\lambda\beta m^2}{2\pi^2a^3}}\right)}\,.\notag
\end{align}
It follows that $\lim_{\beta\to\infty}G_2(\lambda,\beta)=0$ for all $\lambda\in\mathbb{R}$ and $\lim_{\lambda\to\infty}G_1(\lambda,\beta)=0$ for all $\beta>0$. For $\lambda\ge0$ we also have $\lim_{\beta\to0^+}G_2(\lambda,\beta)=\infty$, because for every $\epsilon>0$ we can choose $n\ge0$ large enough and $\beta>0$ small enough to make $\exp\left(\frac{\lambda m^2}{2\pi^2a^2l_n}+\frac{\beta}{a}l_n\right)-1<\epsilon$. Similarly, for all $\beta>0$,
$\lim_{\lambda\to\lambda_0(\beta)^+}G_1(\lambda,\beta)=\infty$. By the Poincar\'e-Miranda theorem \cite{Kulpa1997}, or rather a variation of it with curved boundaries, we can then find values $\beta>0$ and $\lambda>\lambda_0(\beta)$  that solve the equations (\ref{Eqn_Semicl2v2},\ref{Eqn_Semicl1v2}).


When $m=0$ the situation simplifies somewhat. We may assume that $Y_2>Y_1=0$, otherwise there would be at most one solution. We note that $\lambda$ is arbitrary, but we still need to choose $\beta>0$ to solve (\ref{Eqn_Semicl1v2}). The sum
\begin{align}
\tilde{G}_2(\beta)&:=\sum_{n=0}^{\infty}\frac{l_n^2}{a^2}\frac{(n+1)^2}{2\pi^2a^2l_n}\frac{1}{\exp\left(\frac{\beta}{a}l_n\right)-1}\notag
\end{align}
is a continuous function on $\beta>0$ which diverges as $\beta\to0^+$ and which vanishes when $\beta\to\infty$. By the Intermediate Value Theorem we can find a $\beta>0$ such that $\tilde{G}_2(\beta)=Y_2$, which gives a solution. Furthermore, the form of the coefficients coincides with that of a $\beta$-KMS state, cf.~(\ref{eqn_thermalAn}). 
\end{proof*}

\section{Discussion}\label{sec:discussion}

Systems in semi-classical gravity, like the Einstein-Klein-Gordon system that we studied here, typically involve renormalisation parameters that cannot be determined without further input, either from observations or from an underlying theory of quantum gravity. This leads to a complicated situation, where the set of solutions depends on external parameters, which are arbitrary as far as the mathematical structure of the equations is concerned. Nevertheless, we have seen that it is possible to prove some general results concerning the set of solutions, at least when assuming a large amount of symmetry on the two-point distributions of the states and on the spacetime. Moreover, these general conclusions remain valid if one considers solutions in the reduction of order scheme, whose purpose is to remove (at least some of the) spurious solutions.

In particular, we have seen that the system has a unique solution with a symmetric two-point distribution if and only if this solution is the ground state. It would be interesting to know whether this special role of the ground state persists when allowing solutions without symmetry. The general validity of Kay's Theorem is a positive indication, but a more detailed investigation will be necessary to settle this question for general states in an  Einstein static universe, or even in general static spacetimes.

The methods that we used to analyse symmetric states of the semi-classical Einstein-Klein-Gordon system in an Einstein static universe can in principle be generalised to higher dimensions. Although the renormalisation freedom in general spacetimes becomes more complicated, the symmetries reduce it to a perfect fluid form which can still be handled. The generalisation of the group theoretic methods is readily available. The generalisation to open Einstein static universes, where the Cauchy surface has a constant negative curvature, could be more challenging, because the symmetry group becomes non-compact. The compactness of the Cauchy surface also ensured that all solutions can be written in terms of density matrices in the ground state representation, which enabled us to minimise the von Neumann entropy.

Our static solutions may also provide a starting point for investigations into the dynamics of a closed FLRW universe, along the lines of \cite{Pinamonti2011,Gottschalk+2018}. Furthermore, it would be of interest to investigate the fluctuations of the components of the renormalised stress tensor and to compare their (relative) size to the fluctuations in the Minkowski vacuum. This would provide an indication whether the solutions that we have found can be viewed as reliable approximations of a physical state in quantum gravity.

\section*{Acknowledgements}

I would not have considered the reduction of order scheme if Bernard Kay hadn't suggested it to me. I thank him for this and other suggestions. I am also grateful for comments by H.~Gottschalk, T.-P.~Hack, D.~Siemssen and P.~Taylor and to the participants of the conference ``The Semi-Classical Einstein Equation: Numerical and Analytical Challenges'' in Dublin (2019), which was funded by the IRC under the New Foundations scheme. Finally I would like to thank two anonymous referees for their careful reading of the manuscript and their comments.

\appendix

\section{The total classical energy}\label{App_TotalE}

\begin{lemma}
The two definitions of total energy of a classical solution $\varphi$ of the Klein-Gordon equation with space-like compact support, given in (\ref{Eqn_totalE}) and (\ref{Eqn_totalE2}), coincide, i.e.
\begin{align}
\int_{\Sigma}n^av^bT_{ab}&=\int_{\Sigma}n^av^b\tilde{T}_{ab}
\end{align}
for the tensors defined in (\ref{Eqn_T}) and (\ref{Eqn_simpleT}) and for every smooth space-like Cauchy surface $\Sigma$.
\end{lemma}
\begin{proof*}
We consider the difference $X_{ab}:=T_{ab}-\tilde{T}_{ab}=\xi(\varphi^2R_{ab}-\nabla_a\nabla_b\varphi^2+g_{ab}\Box\varphi^2)$ and note that
\begin{align}
v^bX_{ab}&=\xi(-\varphi^2\Box v_a-v^b\nabla_b\nabla_a\varphi^2+v_a\Box\varphi^2)\\
&=\xi\nabla^b(-\varphi^2\nabla_bv_a-v_b\nabla_a\varphi^2+v_a\nabla_b\varphi^2)\,.\label{eqn:divergence}
\end{align}
Here we used the fact that $v^a$ is a Killing vector field, so that $\nabla_bv^b=0$ and $\Box v_a=-R_{ab}v^b$ (cf.\ \cite{Wald}, Eqn.\ (C.3.6)). Near the Cauchy surface $\Sigma$ we may introduce Gaussian normal coordinates so that the future pointing normal vector field $n^a$ is extended to a coordinate vector field and $K_{ab}=\nabla_an_b$ is symmetric. We may then introduce
\begin{align}
w_b&:=n^a(-\varphi^2\nabla_bc_a-v_b\nabla_a\varphi^2+v_a\nabla_b\varphi^2)
\end{align}
and because the expression in brackets is anti-symmetric we find from (\ref{eqn:divergence}) that
$n^av^bX_{ab}=\xi\nabla^bw_b$. Note that $w^b$ is tangent to $\Sigma$ and $\nabla_bw^b=\tensor*[^{(\Sigma)}]{\nabla}{_i}w^i$, where the index $i$ refers to components tangent to $\Sigma$ and $\tensor*[^{(\Sigma)}]{\nabla}{}$ is the connection for the metric on $\Sigma$ obtained by restricting $g_{ab}$. Finally, because $\varphi$ has space-like compact support, $w^i$ has compact support on $\Sigma$ and by Stokes' theorem
\begin{align}
\int_{\Sigma}n^av^bX_{ab}=\xi\int_{\Sigma}\tensor*[^{(\Sigma)}]{\nabla}{_i}w^i=0\,.
\end{align}
\end{proof*}

The lemma implies in particular that Equation (\ref{Eqn_totalE2}) is independent of the choice of Cauchy surface. This can also be established directly by showing that $\nabla^av^b\tilde{T}_{ab}=v^b\nabla^a\tilde{T}_{ab}=0$, because
$\nabla^a\tilde{T}_{ab}=-\nabla^aX_{ab}=-\frac12\xi\varphi^2\nabla_bR$ and $v^b\nabla_bR=0$.

\section{Higher spherical harmonics}\label{App_Harmonics}

In this appendix we review some basic facts on higher spherical harmonics that are needed in the main text. Our presentation is based on \cite{Dieudonne1980} (see also \cite{Frye+2012}).

\subsection{Harmonic functions on $\mathbb{S}^p$}\label{appsubsec:harmonic}

For $p\ge1$ we consider the round unit sphere $\mathbb{S}^p$ as a subset of the Euclidean space $\mathbb{R}^{p+1}$ with the induced metric. We denote the Laplace operator on $\mathbb{R}^{p+1}$ by $\Delta=\sum_{j=1}^{p+1}\partial_{x^j}^2$ in Cartesian coordinates and we denote the Laplace-Beltrami operator on $\mathbb{S}^p$ by $\Delta_{\mathbb{S}^p}$. The two operators may be related by introducing spherical coordinates $(r,\theta_1,\ldots,\theta_p)$
in which we have
\begin{align}\label{eqn_sphericalLaplace}
\Delta&=r^{-p}\partial_rr^p\partial_r+r^{-2}\Delta_{\mathbb{S}^p}\,.
\end{align}
Due to the compactness of $\mathbb{S}^p$ and the ellipticity of $-\Delta_{\mathbb{S}^p}$, the Hilbert space $L^2(\mathbb{S}^p)$ (integrating with respect to the natural volume form) has an orthonormal basis of eigenfunctions for $-\Delta_{\mathbb{S}^p}$.
We will review some facts on the eigenvalues and eigenspaces of $-\Delta_{\mathbb{S}^p}$.

A \emph{spherical harmonic} (of degree $n\ge0$) is the restriction $f|_{\mathbb{S}^p}$ of a complex polynomial (of degree $n$) in $\mathbb{R}^{p+1}$ which is harmonic, i.e.~a polynomial satisfying the Laplace equation $\Delta f=0$. We let $\mathscr{H}(n,p+1)$ denote the homogeneous harmonic polynomials of degree $n$ in $\mathbb{R}^{p+1}$ and we denote the space of spherical harmonics of degree $n\ge0$ by $L_n^{(p)}$, which is a subspace of $L^2(\mathbb{S}^p)$. The restriction map $f\mapsto f|_{\mathbb{S}^{p}}$ from $\mathscr{H}(n,p+1)$ to $L_n^{(p)}$ is injective, because $f(x)=r^nf|_{\mathbb{S}^p}\left(\frac{x}{r}\right)$ can be reconstructed from its restriction. Hence, $\mathscr{H}(n,p+1)$ and $L_n^{(p)}$ have the same dimension.

A harmonic polynomial $f$ is uniquely determined by the data $f|_{x^{p+1}=0}$ and $\partial_{x^{p+1}}f|_{x^{p+1}=0}$, which can be chosen freely.
If $f$ is homogeneous of degree $n$, the data are homogeneous polynomials of degree $n$ and $ n-1$, respectively. It follows that the dimension of $L_n^{(p)}$ (and of $\mathscr{H}(n,p+1)$) is given by
\begin{align}
\mathrm{dim}(L_n^{(p)})&=\begin{cases}
1&\mathrm{if}\ n=0\,,\\
\binom{n+p-1}{n}+\binom{n-1+p-1}{n-1} = \frac{2n+p-1}{n+p-1}\binom{n+p-1}{n}&\mathrm{if}\  n\ge 1\,.\label{eqn:dimLnp}
\end{cases}
\end{align}
If $h\in L_n^{(p)}$, then $h=f|_{\mathbb{S}^p}$ for some $f\in\mathcal{H}(n,p+1)$ and we deduce from $\Delta f=0$ and  (\ref{eqn_sphericalLaplace}) that
\begin{align}
-\Delta_{\mathbb{S}^p}h&=n(n+p-1)h\,,
\end{align}
which shows that $L_n^{(p)}$ is an eigenspace for $-\Delta_{\mathbb{S}^p}$ with eigenvalue $\lambda_n^{(p)}=n(n+p-1)$.

We now want to argue that the spherical harmonics generate a dense subset of $L^2(\mathbb{S}^p)$, so the eigenspaces $L_n^{(p)}$ with eigenvalues $\lambda_n^{(p)}$ determine the full spectrum of $-\Delta_{\mathbb{S}^p}$. By the Stone-Weierstrass theorem, every continuous function on $\mathbb{S}^p$ can be approximated uniformly by restrictions of polynomials from $\mathbb{R}^{p+1}$ to $\mathbb{S}^p$. In particular, restrictions of polynomials are dense in $L^2(\mathbb{S}^p)$ and it remains to show that it suffices to consider harmonic polynomials. This follows from the fact that every polynomial of degree $n\ge0$ can be written in the form
\begin{align}\label{Eqn_harmonicdecomposition}
f&=\sum_{m=0}^{\lfloor\frac{n}{2}\rfloor}r^{2m}h_m\,,
\end{align}
where the $h_m$ are harmonic poplynomials of degree $n-2m$. Restricting to $\mathbb{S}^p$ then gives $f|_{\mathbb{S}^p}=\sum_{m=0}^{\lfloor\frac{n}{2}\rfloor}h_m|_{\mathbb{S}^p}$, so the restrictions of harmonic polynomials generate a dense subspace of $L^2(\mathbb{S}^p)$. To see why (\ref{Eqn_harmonicdecomposition}) holds it clearly suffices to consider homogeneous polynomials $f$. When $n=0$ or $n=1$, $f$ is harmonic itself. We then proceed by induction and assume that the decomposition holds for homogeneous polynomials of degree $n$ and that $f$ has degree $n+2$. Because $\Delta f$ has degree $n$ we can write $\Delta f=\sum_{m=0}^{\lfloor\frac{n}{2}\rfloor}r^{2m}\tilde{h}_m$ for some harmonic polynomials $\tilde{h}_m$ which must be homogeneous of degree $n-2m$. Using Euler's equation $\sum_{i=1}^{p+1}x^i\partial_{x^i}\tilde{h}_m=(n-2m)\tilde{h}_m$ one can compute that $h_{m+1}:=\frac{1}{2(m+1)(2n-2m+p+1}\tilde{h}_m$ satisfies $\Delta(r^{2m+2}h_{m+1})=r^{2m}\tilde{h}_m$ and hence $h_0:=f-\sum_{m=1}^{\lfloor\frac{n}{2}\rfloor}r^{2m}h_m$ is harmonic, as desired.

\subsection{Representations of $SO(p+1,\mathbb{R})$}\label{SSec_reps}

The connected Lie group $SO(p+1,\mathbb{R})$ acts on $\mathbb{S}^p$ by restricting its defining representation on $\mathbb{R}^{p+1}$. We write $x_N=(0,\ldots,0,1)\in\mathbb{S}^p$ (the ``north pole'') and we note that for $p\ge 2$ the embedding
$\mathbb{R}^p\simeq \mathbb{R}^p\times\{0\}\subset\mathbb{R}^{p+1}$ leads to an embedding $\mathbb{S}^{p-1}\subset\mathbb{S}^p$ (the ``equator''). We may identify $SO(p,\mathbb{R})$ with the subgroup of $SO(p+1,\mathbb{R})$ which leaves $x_N$ invariant and
we may identify the sphere as the homogeneous space $\mathbb{S}^p=SO(p+1,\mathbb{R})/SO(p,\mathbb{R})$.

Because the action of $SO(p+1,\mathbb{R})$ on $\mathbb{R}^{p+1}$ preserves homogeneous polynomials and because it commutes with $\Delta$, it restricts to an action on the space $\mathscr{H}(n,p+1)$ and hence also on the space $L_n^{(p)}$. For any $g\in SO(p+1,\mathbb{R})$ we write $U_n^{(p)}(g)$ for the unitary operator on $L_n^{(p)}$ that implements $g$.

In the special case $p=1$, the spaces $\mathscr{H}(n,p+1)$, $n\ge1$, are two-dimensional and generated by the polynomials $(x+iy)^n$ and $(x-iy)^n$, which restrict to $e^{in\varphi}$ and $e^{-in\varphi}$, using the notation $x+iy=re^{i\varphi}$. Each of these basis functions spans a one-dimensional representation of the group $SO(2,\mathbb{R})$, where rotation over an angle $\theta$ acts as multiplication by $e^{i\theta}$.

For $p\ge 2$ we now want to show that $U_n^{(p)}$ defines an irreducible representation. For $n=0$ this is obvious on dimensional grounds, so we may assume $n\ge 1$. Let $V\subset L_n^{(p)}$ be any non-trivial linear subspace. We can find an orthonormal basis $\{h_1,\ldots,h_d\}$ for $V$ with some $1\le d\le \mathrm{dim}(L_n^{(p)})$. We may write the orthogonal projection $E_V$ onto $V$ as an integral kernel in $C^{\infty}(\mathbb{S}^p\times\mathbb{S}^p)$, namely
\begin{align}
E_V(x,x')&=\sum_{j=1}^dh_j(x)\overline{h_j(x')}\,.\label{eqn:EV}
\end{align}
If $V$ is invariant under the action of $SO(p+1,\mathbb{R})$, then $E_V$ commutes with the representation $U_n^{(p)}$ on $L_n^{(p)}$, which means that $E_V(g\cdot  x,x')=E_V(x,g^{-1}\cdot x')$ for all $g\in SO(p+1,\mathbb{R})$. It follows that $E_V(x,x')$ is uniquely determined by the function $e_V(x):=E_V(x,x_N)$, which is invariant under the action of the subgroup $SO(p,\mathbb{R})$ which leaves $x_N$ invariant. Note that $e_V(x)=\sum_{j=1}^dh_j(x)\overline{h_j(x_N)}$ is itself in $L_n^{(p)}$ (and even in the subspace $V$). Let $f\in \mathscr{H}(n,p+1)$ be a harmonic polynomial such that
\begin{align}
e_V&=f|_{\mathbb{S}^p}\,.\label{eqn:eVrestr}
\end{align}
Recall that $f$ is uniquely determined by the data $f|_{x^{p+1}=0}$ and $\partial_{x^{p+1}}f|_{x^{p+1}=0}$, which must be homogeneous polynomials of degree $n$ and $n-1$, respectively. Moreover, these data must be invariant under $SO(p,\mathbb{R})$. When $p\ge 2$
it follows that $f_n(x)=c_nr^n$ and $f_{n-1}(x)=c_{n-1}r^{n-1}$ for some constants $c_n$ and $c_{n-1}$. Since $f_n$ and $f_{n-1}$ are polynomials, $r$ must have an even power, i.e.~$c_n=0$ when $n$ is odd and $c_{n-1}=0$ when $n$ is even. These conditions uniquely determine $f(x)$ up to scale, and hence $e_V(x)$ is also uniquely determined up to scale. The scale is fixed by requiring $E_V^2=E_V$. This means that there can only be one projection $E_V$ onto a non-trivial invariant subspace. Therefore,
$V=L_n^{(p)}$ and $U_n^{(p)}$ is an irreducible representation for $p\ge 2$.

\subsection{Gegenbauer polynomials}\label{SSec_Gegenbauer}

In the notation of subsection \ref{SSec_reps} we now consider the orthogonal projection $E_n^{(p)}$ onto the eigenspace
$L_n^{(p)}$ of $-\Delta_{\mathbb{S}^p}$ with eigenvalue $\lambda_n^{(p)}=n(n+p-1)$.  For $p=1$ we have an explicit orthonormal eigenbasis $\{\frac{1}{\sqrt{2\pi}}e^{in\varphi}\}_{n\in\mathbb{Z}}$ with eigenvalues $\lambda_n^{(1)}=n^2$
and hence
\begin{align}
E_n^{(p)}(\varphi,\varphi')&=\begin{cases}
\frac{1}{2\pi}&\mathrm{if}\ n=0\,,\\
\frac{1}{\pi}\cos(n(\varphi-\varphi'))&\mathrm{if}\  n\ge 1\,.\notag
\end{cases}
\end{align}

For $p\ge2$ we recall from (\ref{eqn:EV}) that $E_n^{(p)}(x,x')$ is still a smooth and rotation invariant function, so it is uniquely determined by $e_n^{(p)}(x):=E_n^{(p)}(x,x_N)$, which is a spherical harmonic of degree $n$. For $n=0$ we have
\begin{align}
E_0^{(p)}&\equiv \frac{1}{\Omega_p}=\frac{\Gamma\left(\frac{p+1}{2}\right)}{2\pi^{\frac{p+1}{2}}}\,,\label{eqn:E0}
\end{align}
with $\Omega_p$ the volume (or rather the area) of $\mathbb{S}^p$. For $n\ge1$ we can write $e_n^{(p)}(x)=P_n^{(p)}(\cos(\theta_p))$ in spherical coordinates, because $e_n^{(p)}(x)$ is invariant under the subgroup $SO(p,\mathbb{R})$ which leaves $x_N$ invariant. Hence,
\begin{align}
0&=(\Delta_{\mathbb{S}^p}+\lambda_n^{(p)})e_n^{(p)}\notag\\
&=\left(\frac{1}{\sin^{p-1}(\theta_p)}\partial_{\theta_p}\sin^{p-1}(\theta_p)\partial_{\theta_p}+n(n+p-1)\right)
P_n^{(p)}(\cos(\theta_p))\notag\\
&=\left((1-y^2)\partial_y^2-py\partial_y+n(n+p-1)\right)P_n^{(p)}(y)\,,\notag
\end{align}
where we introduced $y=\cos(\theta_p)$. The last line shows that $P_n^{(p)}$ solves Gegenbauer's differential equation. Up to scale, the unique polynomial solution to this equation is the Gegenbauer polynomial $C_n^{(\frac12(p-1))}(y)$,
which is given by Rodrigues' formula
\begin{align}
C_n^{(\frac12(p-1))}(y)&=c_n^{(p)}(1-y^2)^{1-\frac{p}{2}}\partial_y^n(1-y^2)^{n+\frac{p}{2}-1}\,,\label{eqn:Rodrigues}
\end{align}
where the normalisation constant is
\begin{align}
c_n^{(p)}&=\frac{(-1)^n\Gamma\left(\frac{p}{2}\right)}{2^n\Gamma\left(n+\frac{p}{2}\right)}\binom{n+p-2}{n}\,.\notag
\end{align}
At $y=1$ we then have
\begin{align}
C_n^{(\frac12(p-1))}(1)&=c_n^{(p)}(1+y)^n(1-y)^{1-\frac{p}{2}}\partial_y^n(1-y)^{n+\frac{p}{2}-1}|_{y=1}
=\binom{n+p-2}{n}\,.\label{eqn:Gegenbauer1}
\end{align}
On the other hand, because the dimension of $L_n^{(p)}$ is the trace of $E_n^{(p)}$ we find
\begin{align}
\mathrm{dim}(L_n^{(p)})&=\int_{\mathbb{S}^p}E_n^{(p)}(x,x)\ \mathrm{dvol}_{\mathbb{S}^p}(x)
=E_n^{(p)}(x_N,x_N)\Omega_p=e_n^{(p)}(x_N)\Omega_p\,.\nonumber
\end{align}
Using (\ref{eqn:dimLnp}) and comparing $e_n^{(p)}(x)=P_n^{(p)}(\cos(\theta_p))$ with $C_n^{(\frac12(p-1))}(y)$ we then find
\begin{align}
e_n^{(p)}(x)&=\frac{2n+p-1}{\Omega_p(p-1)}C_n^{(\frac12(p-1))}(\cos(\theta_p))\notag
\end{align}
for all $n\ge 0$ and $p\ge2$. In terms of the geodesic distance $\chi(x,x')$ on $\mathbb{S}^p$, we can recover the integral kernel for the projection onto $L_n^{(p)}$ for all $n\ge0$ and $p\ge2$ as
\begin{align}
E_n^{(p)}(x,x')&=\frac{2n+p-1}{\Omega_p(p-1)}C_n^{(\frac12(p-1))}(\cos(\chi(x,x')))\notag\\
&=\frac{(2n+p-1)\Gamma\left(\frac{p+1}{2}\right)}{2(p-1)\pi^{\frac{p+1}{2}}}
C_n^{(\frac12(p-1))}(\cos(\chi(x,x')))\,.\label{eqn:projectionkernel}
\end{align}

The generating function for the Gegenbauer polynomials is known to be (\cite{AbramowitzStegun1970} Eqn.~(22.9.3))
\begin{align}
\sum_{n=0}^{\infty}z^nC_n^{(\frac12(p-1))}(y)&=(1-2yz+z^2)^{\frac{1-p}{2}}\notag
\end{align}
for $y\in[-1,1]$, where the series converges for $|z|<1$. We will apply this for $p=3$, where we may use (\ref{eqn:projectionkernel},\ref{eqn:E0}) to write this as
\begin{align}
\sum_{n=0}^{\infty}\frac{z^n}{n+1}E_n^{(3)}(x,x')&=\frac{1}{2\pi^2}(1-2\cos(\chi(x,x')) z+z^2)^{-1}\,,\label{eqn:generating}
\end{align}
a result that can be verified by elementary methods by multiplying both sides with $1-2\cos(\chi(x,x')) z+z^2$ and by using the recursion relation $C_{n+2}^{(1)}(y)=2yC_{n+1}^{(1)}(y)-C_n^{(1)}(y)$ for the Gegenbauer polynomials to manipulate the series on the left-hand side.

\section{Derivation of Equations (\ref{eqn:restrictedgroundstate2}) and (\ref{eqn:restrictedgroundstate3})}\label{app:ground}

We expand the eigenvalues $l_n$ of $\sqrt{A}$ in powers of $(n+1)$,
\begin{align}
l_n&=\sqrt{(n+1)^2+c}=(n+1)\left(1+\frac{c}{2(n+1)^2}-\frac{c^2}{8(n+1)^4}+O((n+1)^{-6})\right)\label{eqn:expln}
\end{align}
from which we find
\begin{align}
\frac{(n+1)^2}{l_n}&=(n+1)\left(1-\frac{c}{2(n+1)^2}+O((n+1)^{-4})\right)\,.\label{eqn:expln-1}
\end{align}
Similarly we expand
\begin{align}
e^{-iT_{\epsilon}(l_n-n-1)}&=1-iT_{\epsilon}(l_n-n-1)-\frac12T_{\epsilon}^2(l_n-n-1)^2+\frac{i}{6}T_{\epsilon}^3(l_n-n-1)^3+\frac{1}{24}T_{\epsilon}^4(l_n-n-1)^4\notag\\
&\quad+O((l_n-n-1)^5)\notag\\
&=1-iT_{\epsilon}\left(\frac{c}{2(n+1)}-\frac{c^2}{8(n+1)^3}\right)
-\frac12T_{\epsilon}^2\left(\frac{c}{2(n+1)}-\frac{c^2}{8(n+1)^3}\right)^2
+\frac{i}{6}T_{\epsilon}^3\left(\frac{c}{2(n+1)}\right)^3\notag\\
&\quad+\frac{1}{24}T_{\epsilon}^4\left(\frac{c}{2(n+1)}\right)^4+O((n+1)^{-5})\notag\\
&=1-\frac{icT_{\epsilon}}{2(n+1)}-\frac{c^2T_{\epsilon}^2}{8(n+1)^2}
+\frac{6ic^2T_{\epsilon}+ic^3T_{\epsilon}^3}{48(n+1)^3}
+\frac{24c^3T_{\epsilon}^2+c^4T_{\epsilon}^4}{384(n+1)^4}+O((n+1)^{-5})\,.\label{eqn:expexp}
\end{align}
Combining (\ref{eqn:expln}), (\ref{eqn:expln-1}) and (\ref{eqn:expexp}) up to order $(n+1)^{-2}$ gives
\begin{align}
\frac{(n+1)^2}{l_n}e^{-iT_{\epsilon}(l_n-n-1)}&=(n+1)-\frac{icT_{\epsilon}}{2}-\frac{4c+c^2T_{\epsilon}^2}{8(n+1)}
+O((n+1)^{-2})\,,\notag\\
(n+1)^2l_ne^{-iT_{\epsilon}(l_n-n-1)}&=
(n+1)^3-\frac{icT_{\epsilon}}{2}(n+1)^2
+\frac{4c-cT_{\epsilon}^2}{8}(n+1)+\frac{-6ic^2T_{\epsilon}+ic^3T_{\epsilon}^3}{48}\notag\\
&\quad +\frac{c^4T_{\epsilon}^4-48c^2}{384(n+1)}+O((n+1)^{-2})\,.\notag
\end{align}
Inserting the first line into Equation (\ref{eqn:restrictedgroundstate}) we find
\begin{align}
\omega^{(\infty)}_2((t,x),(t',x))&=\lim_{\epsilon\to0^+}\sum_{n=0}^{\infty}\frac{1}{4\pi^2a^2}
\left((n+1)-\frac{icT_{\epsilon}}{2}-\frac{4c+c^2T_{\epsilon}^2}{8(n+1)}+O((n+1)^{-2})\right)
e^{-iT_{\epsilon}(n+1)}\,.\label{eqn:appC1}
\end{align}
The term $O((n+1)^{-2})$ is uniformly absolutely summable, even in the limit $\epsilon=0$, so it converges uniformly to a continuous function,
\begin{align}
&\lim_{\epsilon\to0^+}\sum_{n=0}^{\infty}\frac{1}{4\pi^2a^2}
\left(\frac{(n+1)^2}{l_n}e^{-iT_{\epsilon}(l_n-n-1)}-(n+1)+\frac{icT_{\epsilon}}{2}+\frac{4c+c^2T_{\epsilon}^2}{8(n+1)}\right)e^{-iT_{\epsilon}(n+1)}\notag\\
=&\ S_1+O(t-t')\,,\notag
\end{align}
where we introduced
\begin{align}
S_1&:=\sum_{n=0}^{\infty}\frac{1}{4\pi^2a^2}\left(\frac{(n+1)^2}{l_n}-(n+1)+\frac{c}{2(n+1)}\right)\,.\label{eqn:defS1}
\end{align}
Similarly we find, using Equation (\ref{eqn:00}),
\begin{align}
\partial_t\partial_{t'}\omega^{(\infty)}_2((t,x),(t',x))&=
\lim_{\epsilon\to0^+}\sum_{n=0}^{\infty}\frac{1}{4\pi^2a^4}
\left((n+1)^3-\frac{icT_{\epsilon}}{2}(n+1)^2
+\frac{4c-cT_{\epsilon}^2}{8}(n+1)
+\frac{-6ic^2T_{\epsilon}+ic^3T_{\epsilon}^3}{48}\right.\notag\\
&\quad \left.+\frac{c^4T_{\epsilon}^4-48c^2}{384(n+1)}+O((n+1)^{-2})\right)
e^{-iT_{\epsilon}(n+1)}\label{eqn:appC2}
\end{align}
where the term $O((n+1)^{-2})$ now gives rise to the continuous function
\begin{align}
&\lim_{\epsilon\to0^+}\sum_{n=0}^{\infty}\frac{1}{4\pi^2a^4}
\left((n+1)^2l_ne^{-iT_{\epsilon}(l_n-n-1)}-(n+1)^3+\frac{icT_{\epsilon}}{2}(n+1)^2
-\frac{4c-cT_{\epsilon}^2}{8}(n+1)\right.\notag\\
&\left.-\frac{-6ic^2T_{\epsilon}-ic^3T_{\epsilon}^3}{48}-\frac{c^4T_{\epsilon}^4-48c^2}{384(n+1)}\right)e^{-iT_{\epsilon}(n+1)}\notag\\
=&\ S_2+O(t-t')\,,\notag
\end{align}
with
\begin{align}
S_2&:=\sum_{n=0}^{\infty}\frac{1}{4\pi^2a^4}\left((n+1)^2l_n-(n+1)^3-\frac{c}{2}(n+1)+\frac{c^2}{8(n+1)}\right)\,.
\label{eqn:defS2}
\end{align}

To evalute the contributions in (\ref{eqn:appC1}, \ref{eqn:appC2}) with higher orders of $n+1$ we introduce $z:=e^{-iT_{\epsilon}}$, which has $|z|<1$ when $\epsilon>0$. By differentiating and integrating the geometric series we obtain
\begin{align}
\sum_{n=0}^{\infty}z^{n+1}&=\frac{z}{1-z}\notag\\
\sum_{n=0}^{\infty}\frac{1}{n+1}z^{n+1}&=\int_0^z\frac{1}{1-w}\mathrm{d}w=-\log(1-z)\notag\\
\sum_{n=0}^{\infty}(n+1)z^{n+1}&=z\partial_z\frac{z}{1-z}=\frac{z}{(1-z)^2}\label{eqn:series}\\
\sum_{n=0}^{\infty}(n+1)^2z^{n+1}&=z\partial_z\frac{z}{(1-z)^2}=\frac{z(1+z)}{(1-z)^3}\notag\\
\sum_{n=0}^{\infty}(n+1)^3z^{n+1}&=z\partial_z\frac{z(1+z)}{(1-z)^3}=\frac{z(1+4z+z^2)}{(1-z)^4}\,,\notag
\end{align}
where the logarithm has its branch cut along the negative real axis. When substituting $z=e^{-iT_{\epsilon}}$ we note that
\begin{align}
\log\left(1-e^{-iT_{\epsilon}}\right)&=\log(iT_{\epsilon})+O(T_{\epsilon})
=\frac12\log\left(-T_{\epsilon}^2\right)+O(T_{\epsilon})\,,\notag
\end{align}
because $\mathrm{Re}(iT_{\epsilon})>0$. Using the series in (\ref{eqn:series}) we find
\begin{align}
\omega^{(\infty)}_2((t,x),(t',x))&=\lim_{\epsilon\to0^+}\frac{1}{4\pi^2a^2}\left(\frac{z}{(1-z)^2}-\frac{icT_{\epsilon}}{2}\frac{z}{1-z}+\frac{4c+c^2T_{\epsilon}^2}{8}\log(1-z)\right)+S_1+O(t-t')\notag\\
&=\lim_{\epsilon\to0^+}\frac{1}{4\pi^2a^2}\left(\frac{1}{2\cos(T_{\epsilon})-2}-\frac{icT_{\epsilon}}{2}\frac{1}{e^{iT_{\epsilon}}-1}
+\frac{4c+c^2T_{\epsilon}^2}{8}\log\left(1-e^{-iT_{\epsilon}}\right)\right)\notag\\
&\quad +S_1+O(t-t')\notag\\
&=\lim_{\epsilon\to0^+}\frac{1}{4\pi^2a^2}\left(\frac{-1}{T_{\epsilon}^2}-\frac{1}{12}-\frac{c}{2}
+\frac{c}{4}\log(-T_{\epsilon}^2))\right)+S_1+O(t-t')\notag\\
&=\lim_{\epsilon\to0^+}\frac{1}{4\pi^2a^2}\left(\frac{-1}{T_{\epsilon}^2}+\frac{c}{4}\log(-T_{\epsilon}^2)\right)
+\frac{1}{4\pi^2a^2}X_1+O(t-t')\,,\label{eqn:appC3}
\end{align}
where
\begin{align}
X_1&:=4\pi^2a^2S_1-\frac{1}{12}-\frac{c}{2}
=4\pi^2a^2S_1-\frac{1+6c}{12}\,.\label{eqn:defX1}
\end{align}
Similarly we find
\begin{align}
\partial_t\partial_{t'}\omega^{(\infty)}_2((t,x),(t',x))&=\lim_{\epsilon\to0^+}\frac{1}{4\pi^2a^4}\left(
\frac{z(1+4z+z^2)}{(1-z)^4}
-\frac{icT_{\epsilon}z(1+z)}{2(1-z)^3}
+\frac{(4c-cT_{\epsilon}^2)z}{8(1-z)^2}\right.\notag\\
&\quad\left.
+\frac{(-6ic^2T_{\epsilon}+ic^3T_{\epsilon}^3)z}{48(1-z)}
-\frac{c^4T_{\epsilon}^4-48c^2}{384}\log(1-z)\right)+S_2+O(t-t')\notag\\
&=\lim_{\epsilon\to0^+}\frac{1}{4\pi^2a^4}\left(
\frac{4+2\cos(T_{\epsilon})}{(2\cos(T_{\epsilon})-2)^2}
+\frac{cT_{\epsilon}\sin(T_{\epsilon})}{(2\cos(T_{\epsilon})-2)^2}
+\frac{(4c-cT_{\epsilon}^2)}{16(\cos(T_{\epsilon})-1)}\right.\notag\\
&\quad\left.
+\frac{(-6ic^2T_{\epsilon}+ic^3T_{\epsilon}^3)}{48\left(e^{iT_{\epsilon}}-1\right)}
-\frac{c^4T_{\epsilon}^4-48c^2}{384}\log\left(1-e^{-iT_{\epsilon}}\right)\right)+S_2+O(t-t')\notag\\
&=\lim_{\epsilon\to0^+}\frac{1}{4\pi^2a^4}\left(\frac{6}{T_{\epsilon}^4}+\frac{1}{120}+\frac{c}{T_{\epsilon}^2}
-\frac{c}{2T_{\epsilon}^2}+\frac{c}{12}-\frac{c^2}{8}+\frac{c^2}{16}\log(-T_{\epsilon}^2)\right)\notag\\
&\quad+S_2+O(t-t')\notag\\
&=\lim_{\epsilon\to0^+}\frac{1}{4\pi^2a^4}\left(\frac{6}{T_{\epsilon}^4}+\frac{c}{2T_{\epsilon}^2}
+\frac{c^2}{16}\log(-T_{\epsilon}^2)\right)+\frac{1}{4\pi^2a^4}X_2+O(t-t')\,,\label{eqn:appC4}
\end{align}
where
\begin{align}
X_2&:=4\pi^2a^4S_2+\frac{1}{120}+\frac{c}{12}-\frac{c^2}{8}
=4\pi^2a^4S_2+\frac{1+10c-15c^2}{120}\,.\label{eqn:defX2}
\end{align}

Equations (\ref{eqn:appC3}) and (\ref{eqn:appC4}) equal (\ref{eqn:restrictedgroundstate2}) and (\ref{eqn:restrictedgroundstate3}) when we notice that the value of $X_1$ in (\ref{eqn:X1}) is obtained from (\ref{eqn:defX1}) and (\ref{eqn:defS1}) and the value of $X_2$ in (\ref{eqn:X2}) is obtained from (\ref{eqn:defX2}) and (\ref{eqn:defS2}).

\end{document}